\documentclass[11pt]{aastex}
\usepackage{graphicx,emulateapj5,apjfonts}
\usepackage{onecolfloat}
\usepackage{epsf}

\shortauthors{Bell et al.}
\shorttitle{Stellar Population variations in the Galactic Halo}

\slugcomment{{\sc The Astronomical Journal, in press} }

\begin{document}


\def\head{

\title{Stellar Population Variations in the Milky Way's Stellar Halo}

\author{Eric F.\ Bell$^{1}$, Xiang Xiang Xue$^{2,3}$, Hans-Walter Rix$^2$, Christine Ruhland$^2$, David W.\ Hogg$^{4,2}$}
\affil{$^1$ Department of Astronomy, University of Michigan, 
500 Church St., Ann Arbor, MI 48109, USA; ericbell@umich.edu \\
$^2$ Max-Planck-Institut f\"ur Astronomie,
K\"onigstuhl 17, D-69117 Heidelberg, Germany \\
$^3$ Key Laboratory of Optical Astronomy, National Astronomical Observatories, CAS, 20A Datun Road, Chaoyang District, 100012, Beijing, China \\
$^4$ Center for Cosmology and Particle Physics,
Department of Physics,
New York University,
4 Washington Place \#424,
New York, NY 10003, USA}

\begin{abstract}
If the stellar halos of disk galaxies are 
built up from the disruption of 
dwarf galaxies, models predict highly structured variations
in the stellar populations within these halos.  
We test this prediction by studying the ratio
of blue horizontal branch stars (BHB stars; more abundant in old, metal-poor populations) to main-sequence turn-off
stars (MSTO stars; a feature of all populations) in the stellar halo of 
the Milky Way using data from the Sloan Digital Sky Survey.
We develop and apply an improved technique to select BHB
stars using $ugr$ color information alone, yielding a sample
of $\sim$9000 $g <18$ candidates where $\sim70$\% of them are BHB stars.
We map the BHB/MSTO ratio across $\sim1/4$ of the sky 
at the distance resolution permitted by the absolute
magnitude distribution of MSTO stars.  We find
large variations of BHB/MSTO star ratio in the stellar halo. 
Previously identified, stream-like halo structures have 
distinctive BHB/MSTO ratios, indicating different ages/metallicities.  
Some halo features, e.g., 
the low-latitude structure, appear to be almost 
completely devoid of BHB stars, whereas other 
structures appear to be rich in BHB stars.  The 
Sagittarius tidal stream shows an apparent variation in BHB/MSTO 
ratio along its extent, which we interpret in terms of 
population gradients within the progenitor dwarf galaxy.
Our detection of coherent stellar population variations 
between different stellar halo substructures 
provides yet more support to 
cosmologically motivated models 
for stellar halo growth.
\end{abstract}

\keywords{galaxies: evolution --- galaxies: bulges --- 
galaxies: spiral --- galaxies: general ---  
galaxies: stellar content }
}

\twocolumn[\head]

\section{Introduction}

Over the last decade, it has become clear that a sizeable, perhaps dominant, 
part of the
stellar halo of the Milky Way is composed of the accumulated 
debris from the disruption 
of distinct satellite galaxies.  Starting with the discovery and 
characterization of the tidal debris from the disruption of the 
Sagittarius dwarf galaxy \citep{ibata95,majewski03}, a variety
of other stellar halo overdensities have been 
discovered: the low-latitude stream \citep{yanny03,ibata03}, 
the Virgo and Hercules-Aquila overdensities \citep{duffau06,juric08,her_aq}, 
and a number of smaller tidal streams 
\citep{pal5,grillmair_orphan,grillmair_stream,grillmair_n5466,orphan,grillmair08,newberg09}.
Models of stellar halos that formed in a cosmological context 
through the disruption of dwarf galaxies 
have been developed and refined
\citep{bkw,bullock05,abadi06,cooper10}, providing 
quantitative predictions for the structure and 
stellar content of stellar halos formed in such a fashion.
Remarkably, the structure and degree of substructure 
of the model stellar halos appear to be in quantitative agreement with 
observations of the Milky Way's stellar halo 
(\citealp{bullock05}, \citealp{bell08}): 
stellar halos with $\sim 10^9 L_{\sun}$, a 
roughly $r^{-3}$ density profile, and showing 
rich substructure on a variety of spatial scales.

A clear prediction of such a picture is that any population 
differences in the stellar halo, arising from different 
progenitor populations, should have a similar morphology 
to the stellar density inhomogeneities.
The metallicity--mass correlation of satellite galaxies 
will translate into 
variations in stellar halo metallicity, as
debris from larger satellites will have higher metallicity than 
those of lower-luminosity satellites or globular clusters.  
Furthermore, one expects
to see distinctive signatures in age and detailed element
abundance patterns, reflecting when satellites were
accreted (\citealp{robertson05}, \citealp{font06}; see also 
\citealp{tumlinson10}).
Signatures of population inhomogeneity have been detected: 
hints of chemical signatures of kinematically-detected local
streams \citep[e.g.,][]{helmi06}, small variations in metallicity 
within the nearby parts of the stellar halo, especially associated with 
the low-latitude stream \citep[e.g.][]{ivezic08}\footnote{A particular
concern in this case is that the low-latitude stream may 
have a significant or even dominant contribution from the debris
torn off of the Milky Way disk; thus, the detection of a distinctive
population may result from disk debris superimposed on a relatively homogeneous
stellar halo.},
the tendency for the halo to have metal-poorer stars at larger 
radii \citep[e.g.,][]{carollo07}, and a recent color--magnitude diagram
(CMD) fitting analysis of Sloan Digital Sky Survey (SDSS) 
stripes showing both a radial metallicity gradient and 
large density and metallicity
variations around that gradient \citep{dejong10}.  
Yet, for the most part,  
the observational signatures of such population variations 
are accessible for relatively bright stars only (high S/N $ugriz$ 
imaging, or moderate S/N spectra).  Accordingly, 
those radii in our own stellar halo where substructure 
is rich ($10<r<40$\,kpc) are not probed particularly well. 
Some of the most impressive
evidence for stellar halo population variations 
are from M31, where significant 
variations in stellar age and metallicity 
have been documented \citep{brown06,richardson08}.
In particular, \citet{ibata07} and \citet{mcconnachie09} 
have used the color of the red giant branch stars 
(practically inaccessible in the Milky Way's 
halo because distances are uncertain) 
to provide panoramic views
of the stellar populations in the M31 halo, showing clearly that 
stellar population variation can be associated
with particular substructures. 

In this paper, we present an exploration of
stellar population variations and their morphology within the stellar 
halo of the Milky Way.  We make use of 
the ratio of blue horizontal branch (BHB) stars
to main-sequence turn-off (MSTO) stars as our main 
observational diagnostic of stellar 
population variations.  
While the exact conditions under
which BHB stars develop are not completely
clear (post-main sequence-stars with core helium burning and 
hydrogen shell burning, 
with relatively low hydrogen envelope masses; \citealp{hoyle55}),
studies of globular clusters demonstrate that they  
signpost particularly ancient and metal-poor stellar populations.
We take MSTO stars (present in all populations) as a proxy for the 
general stellar content of the halo; thus, variations in BHB to MSTO 
ratio qualitatively probe the ratio of 
ancient and metal poor stars to all stars.  
Despite our lack of a complete 
understanding of which populations harbor BHB stars, these stars are in many 
other respects ideal probes of population variations in the stellar halo:
they can be selected from the SDSS imaging and spectroscopic dataset, 
they are relatively luminous (allowing probing to large distances), and 
they are nearly standard candles.  

We outline and apply an improved color selection method
to isolate a sample of over 9000 $g\la18$ high-probability BHB 
star candidates from SDSS data (\S \ref{sec:data}).  
The MSTO stars are color-selected
in a simple empirical fashion to encompass the full range of MSTO colors
in the stellar halo; we therefore use MSTO stars as a proxy
for the general stellar content of the stellar halo.  
We then explore the relative distributions
of BHB to MSTO stars in the Heliocentric distance range 
$5 \la r/{\rm kpc} \la 30$ across $\sim 1/4$ of the celestial sphere, 
providing a panoramic view of stellar population variations in
the Milky Way's stellar halo (\S \ref{sec:bhbmsto}). 
Owing to our uncertainty in the exact conditions under 
which BHB stars form, such a view of stellar population 
variations is necessarily qualitative in nature, but has the 
advantage that stellar population variations can be traced
out to larger distances than is currently possible with 
complementary tracers (e.g., the lower main sequence stars
of \citealp{ivezic08}).
We then compare qualitatively
with models of stellar halos built entirely through the disruption
of dwarf galaxies (\S \ref{sec:mod}), 
and explore some other implications of our results
(\S \ref{sec:disc}).

\section{BHB and MSTO star samples} \label{sec:data}

\begin{figure}[t]
\begin{center}
\epsfbox{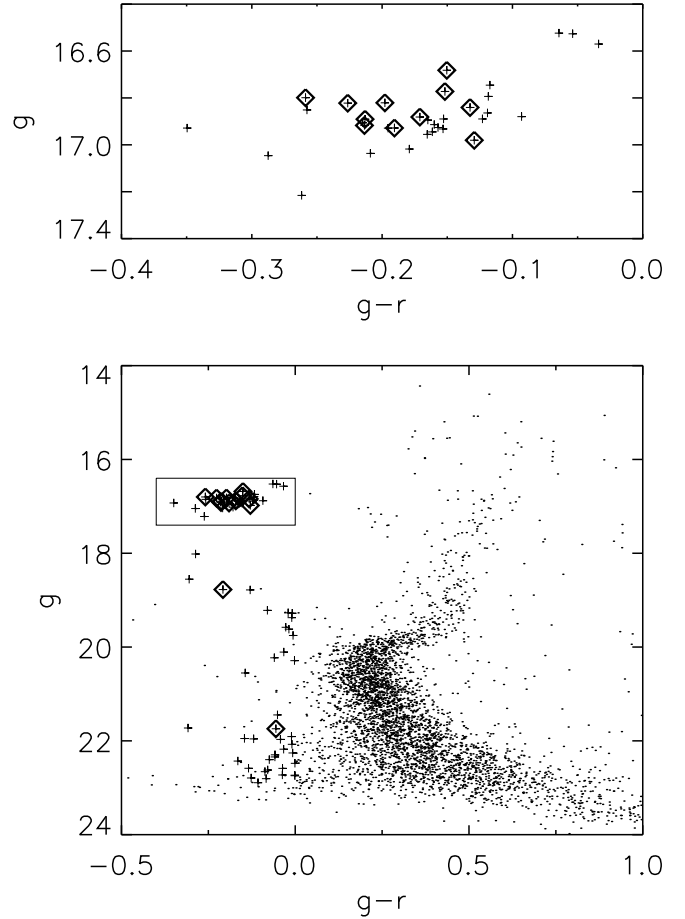}
\end{center}
\caption{\label{fig:gc}
Illustration of the recovery of BHB stars for the globular
cluster NGC 5024.   The bottom panel shows the cluster
CMD over a wide range of colors and magnitudes; 
the top panel zooms into the area richest in BHB stars.  Stars selected
as BHB candidates by $-0.5<g-r<0$ and $0.8<u-g<1.6$
are shown as crosses, while those with $>50\%$ probability of being
a BHB star
as estimated by our method are marked with an additional diamond.   
}
\end{figure}

\begin{figure}[th]
\begin{center}
\epsfbox{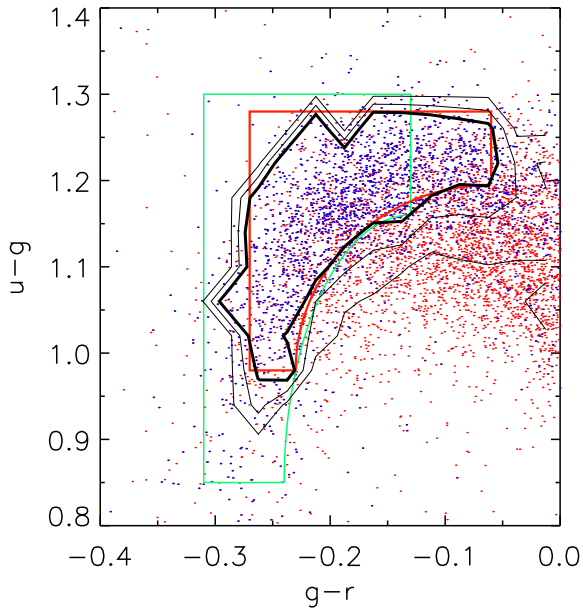}
\end{center}
\caption{\label{fig:selection}
Illustration of the BHB star selection method which identifies
the optimal BHB selection region in the $u-g$ vs.\ $g-r$ color plane using
extensive spectroscopic identification using the SDSS.  
Only stars with spectroscopic 
identifications are shown.  Stars 
that are argued to be BHB stars on the basis of their 
spectra by \protect\citet{xue08} are shown in blue.
Contours show the fraction 
of candidates determined to be BHB stars (contours are 10\%, 30\% and 
50\%; the 50\% contour is thicker than the others).  
In what follows we consider BHB stars within the 50\% contour only.
Overplotted are our approximation to the BHB color selection limits (red), and 
the 'stringent' color cut of \protect\citet[green]{sirko04}.
}
\end{figure}

\begin{figure}[th]
\begin{center}
\includegraphics[width=8.0cm]{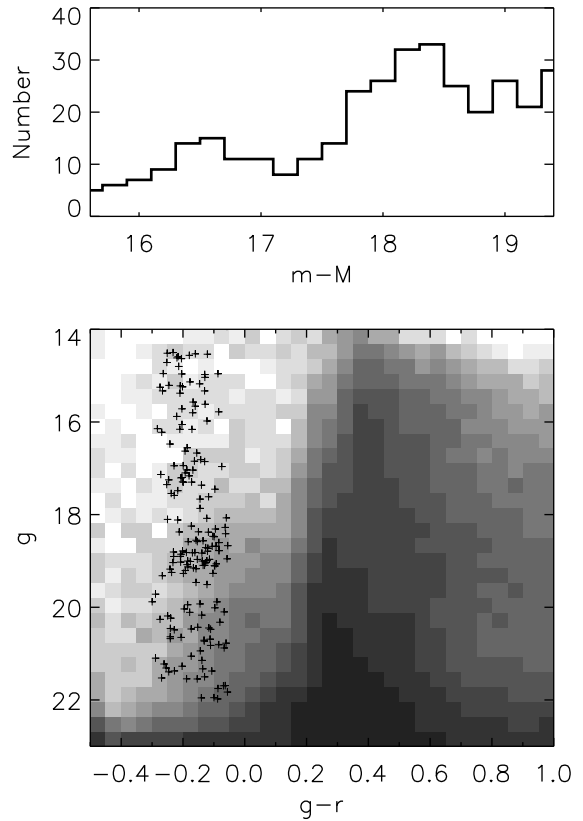}
\end{center}
\caption{\label{fig:sgr}
BHB star candidates in the direction of RA/Dec=(204,5).  The lower
panel shows the logarithmic density of stars in the CMD in gray scale
in an area of radius 4$^{\arcdeg}$, with those stars
identified as BHB candidates on the basis of their $ugr$ colors
being shown with crosses.  The upper panel shows the distance
modulus distribution of BHB candidates along this line of sight, 
showing two overdensities; the more distant of these at $m-M \sim 18.3$ 
is part of the tidal tail from Sagittarius.  It is worth noting that 
blue straggler contamination has a much broader absolute 
magnitude distribution; 
the confinement of this overdensity to such a small apparent magnitude 
range is further evidence that the sample is dominated by BHB stars
(that have a small range in absolute magnitude).
}
\end{figure}

\begin{figure*}[t]
\begin{center}
\includegraphics[width=4.7cm]{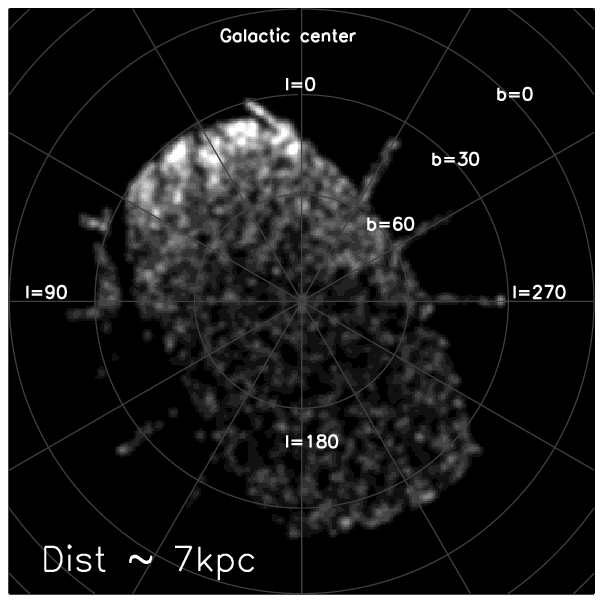}
\includegraphics[width=4.7cm]{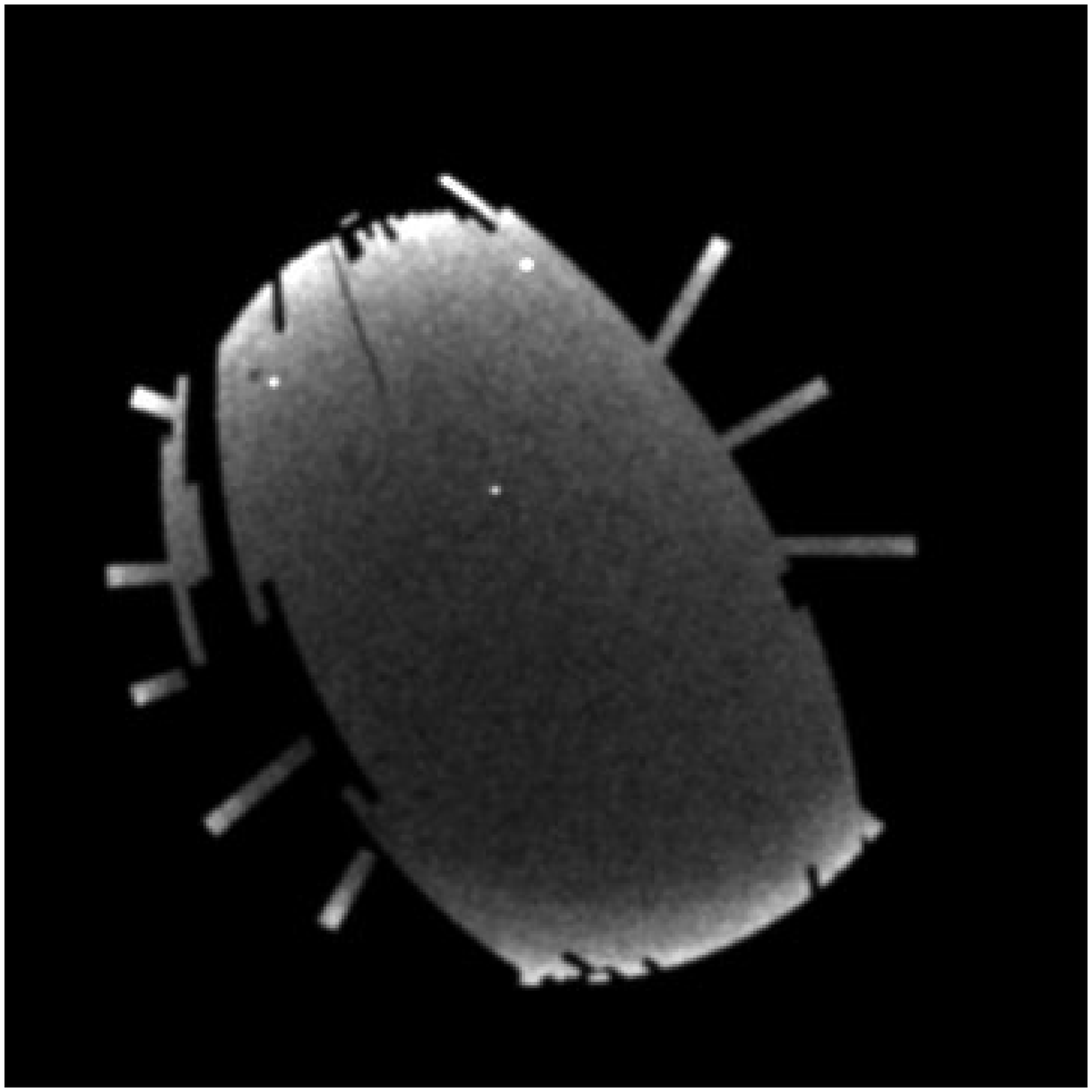}
\includegraphics[width=4.7cm]{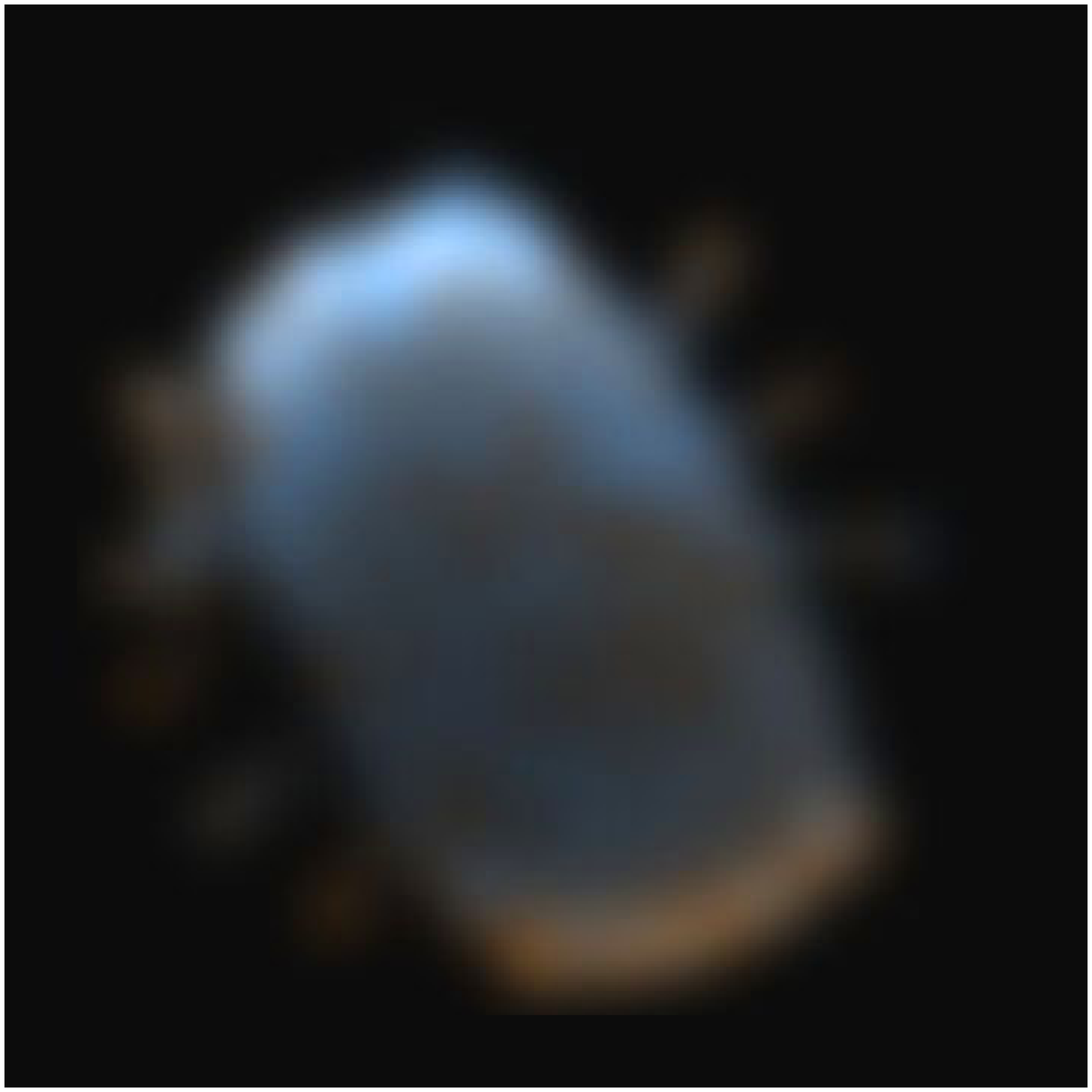}
\includegraphics[width=4.7cm]{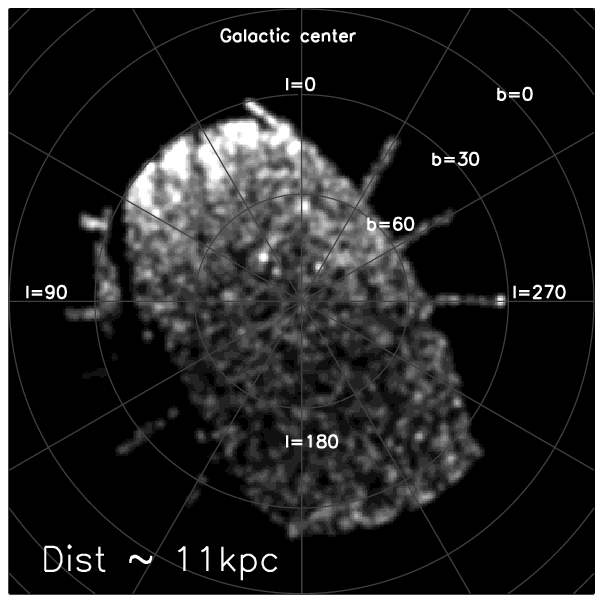}
\includegraphics[width=4.7cm]{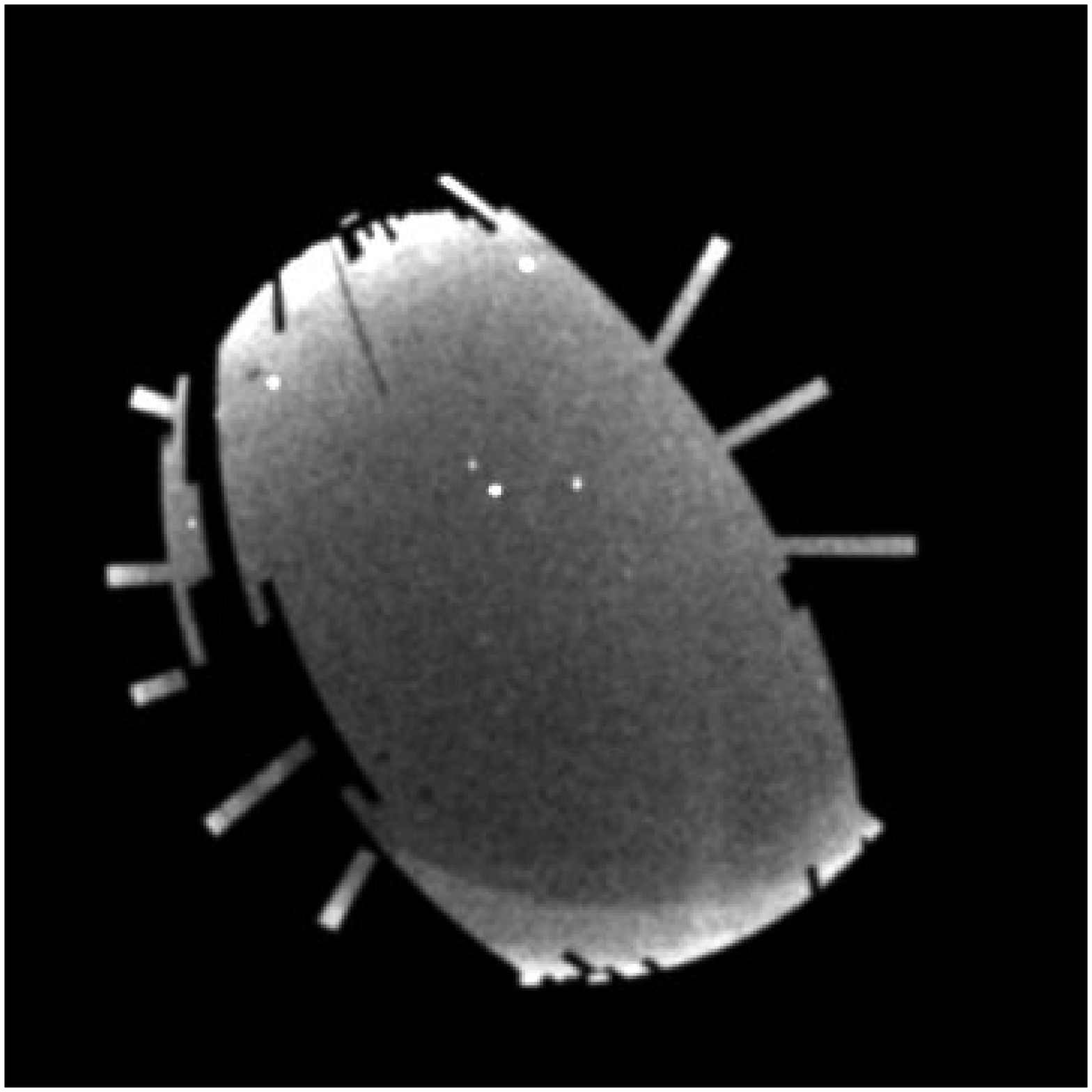}
\includegraphics[width=4.7cm]{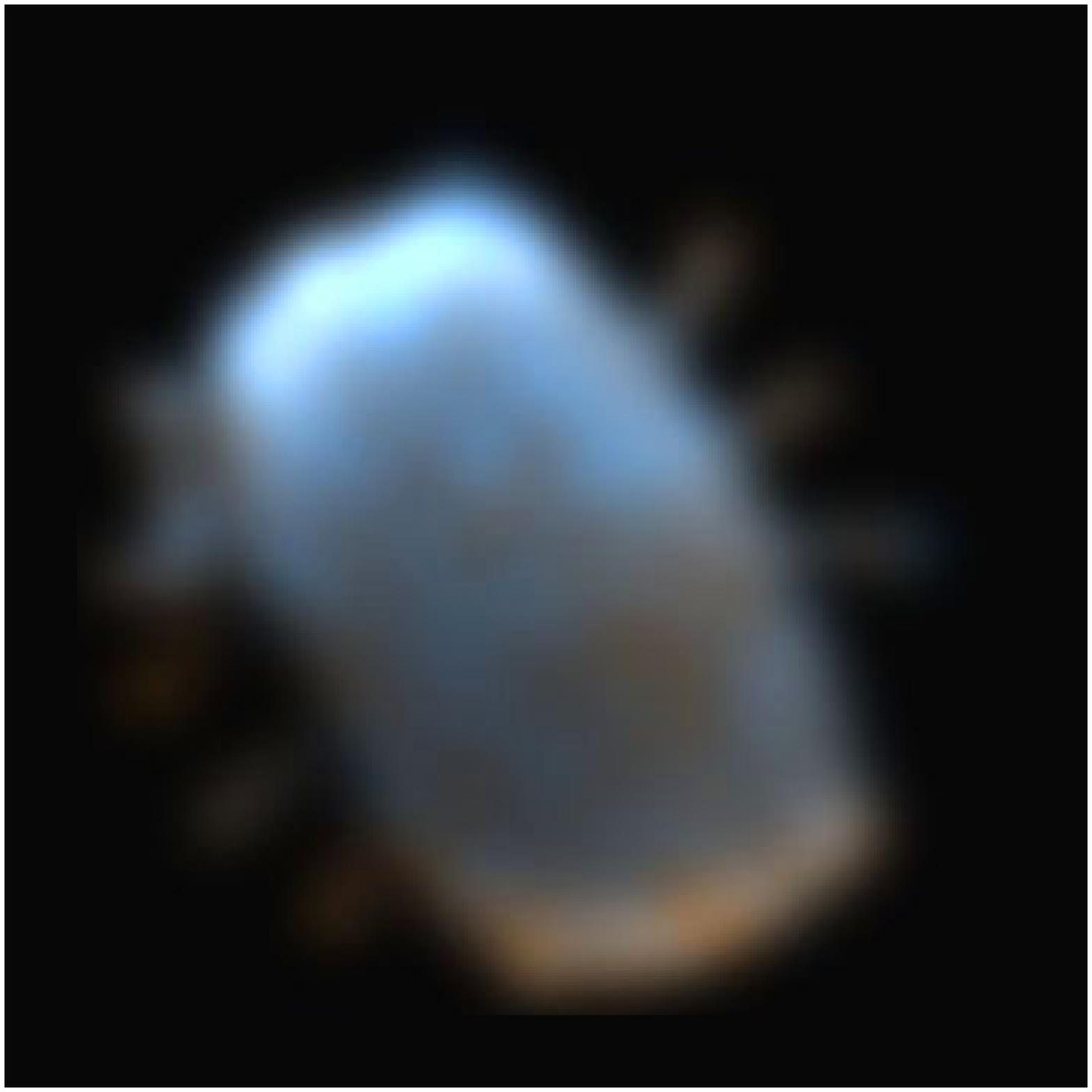}
\includegraphics[width=4.7cm]{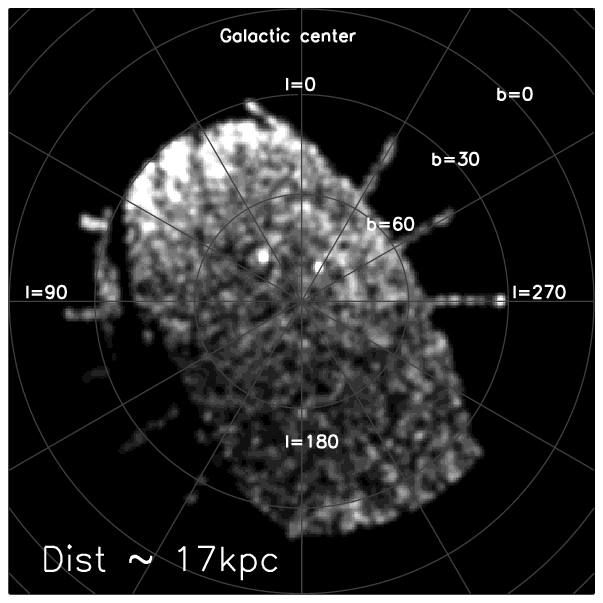}
\includegraphics[width=4.7cm]{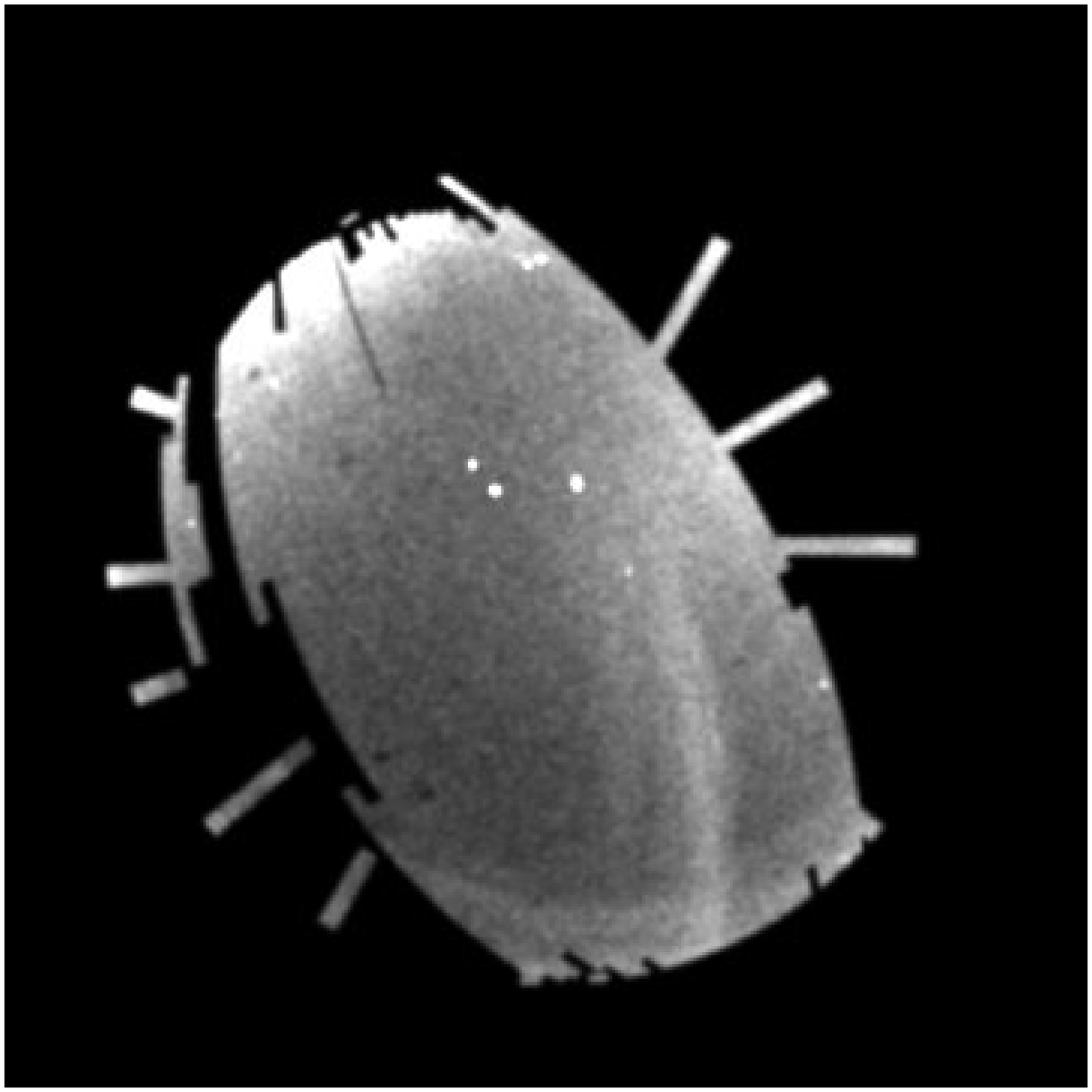}
\includegraphics[width=4.7cm]{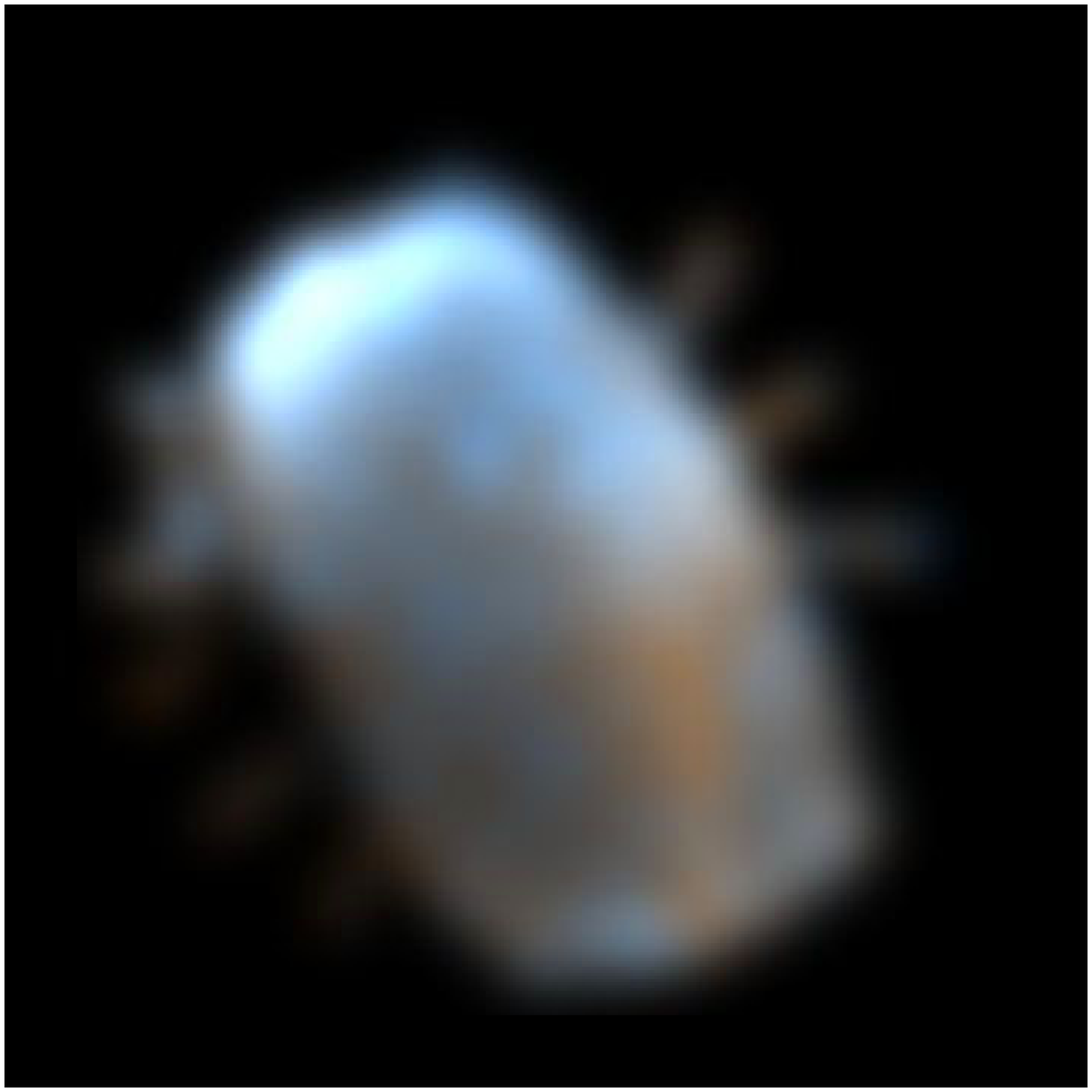}
\includegraphics[width=4.7cm]{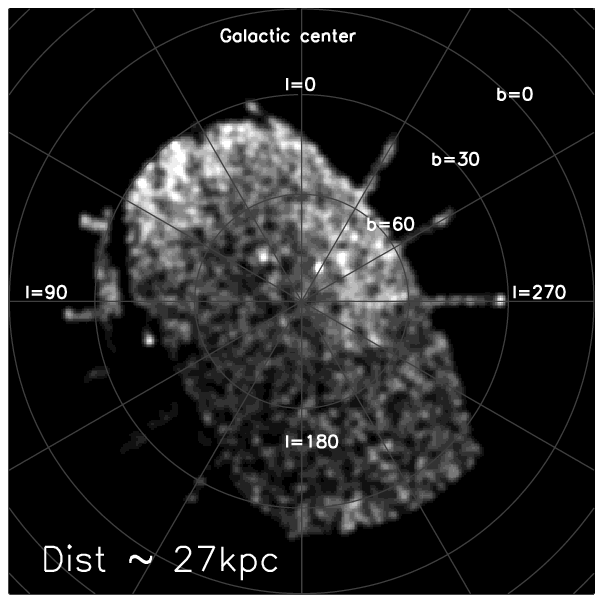}
\includegraphics[width=4.7cm]{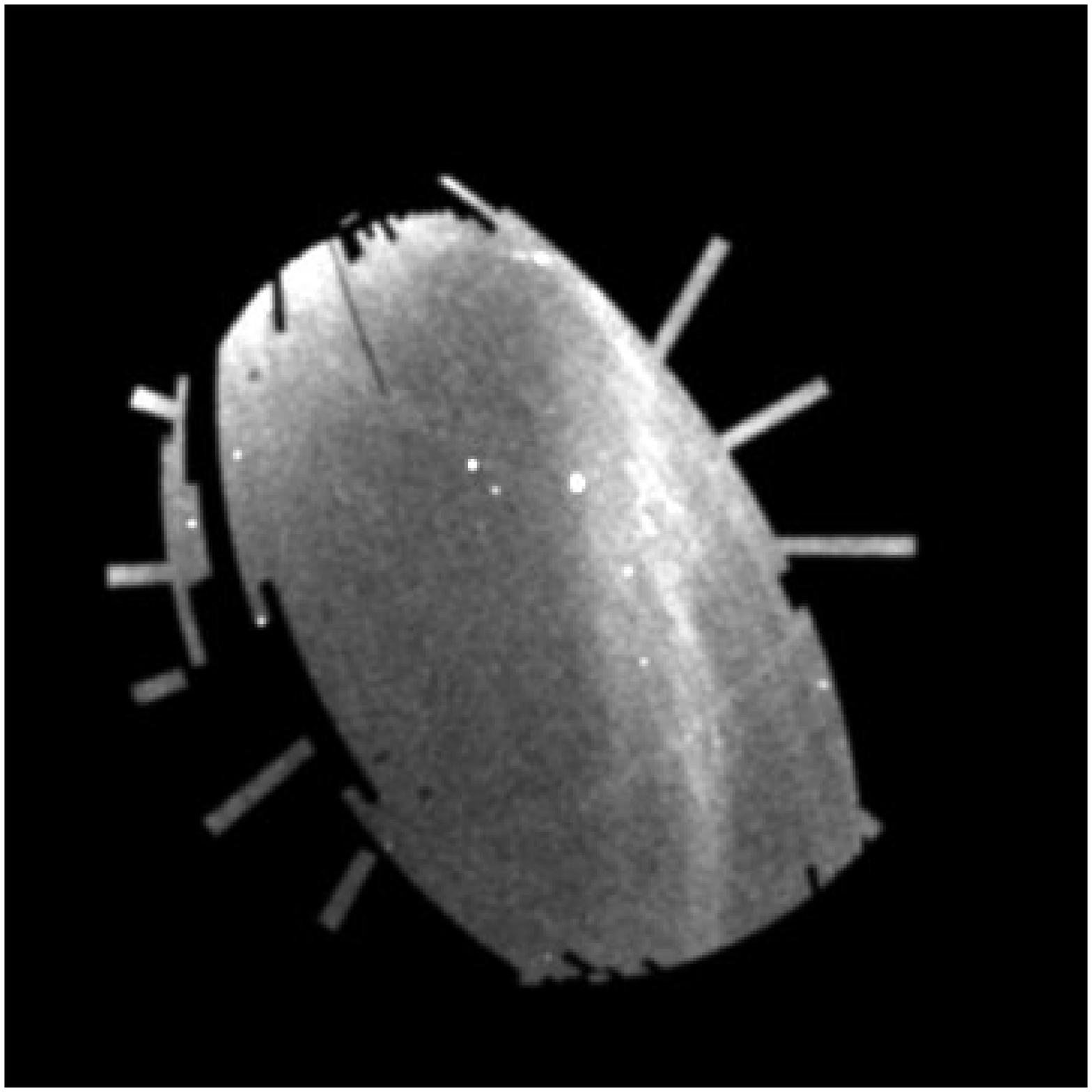}
\includegraphics[width=4.7cm]{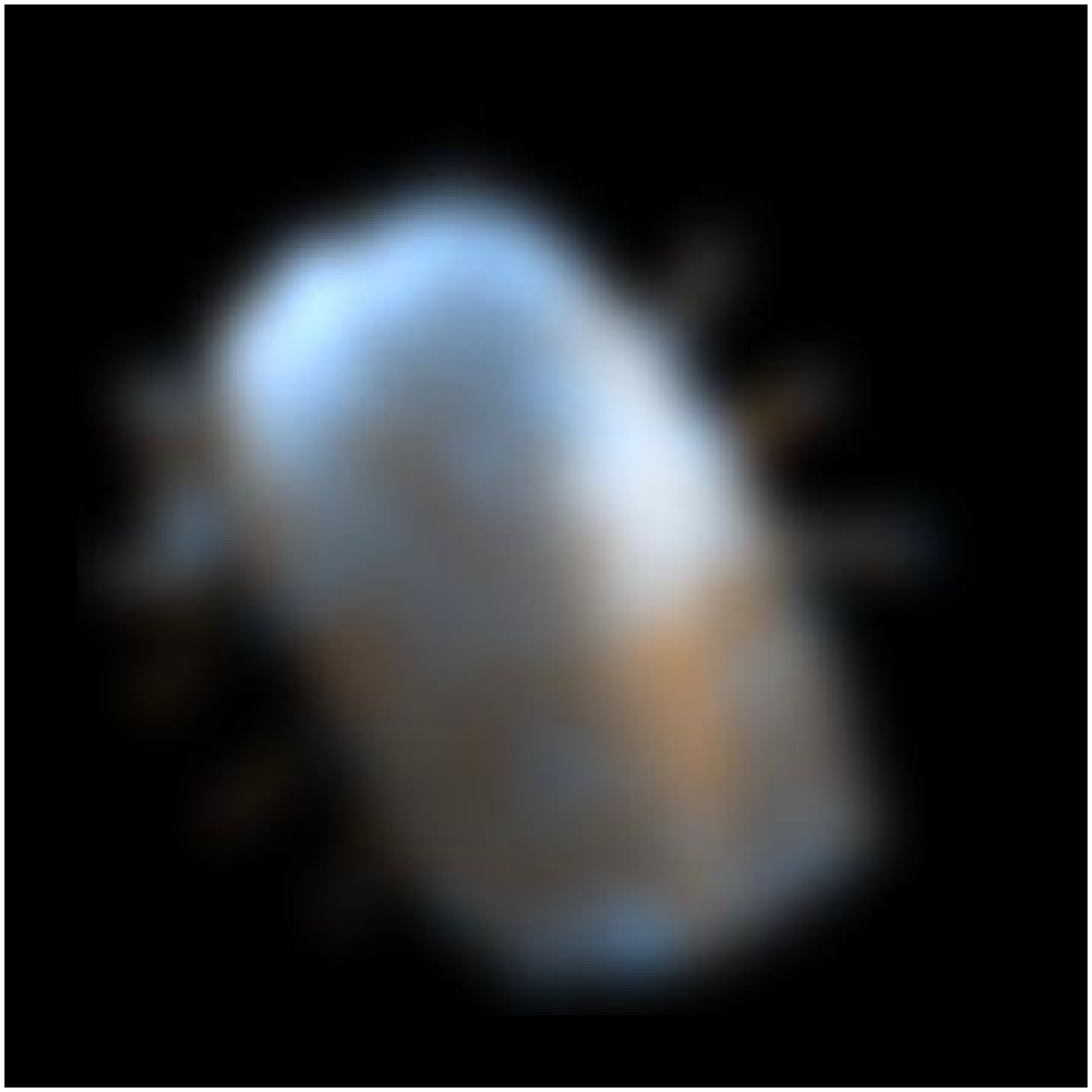}
\end{center}
\caption{\label{fig:map} 
Left: Map of the BHB stars in thick distance modulus slices, where the BHB distances have been `degraded' to $\sigma_{M} = 0.9$ to match the distance resolution of MSTO stars.  Middle: Map of MSTO stars in the same distance modulus slice, assuming a MSTO $M_r = 4.5$.
Right: Color representation of the BHB/MSTO ratio, smoothed with a 6 degree Gaussian.  Panels are arranged in order of increasing distance modulus from 
top to bottom, $13.5 \le m-M < 14.5$, $14.5 \le m-M < 15.5$, 
$15.5 \le m-M < 16.5$, and $16.5 \le m-M < 17.5$, or distances of
$5-8$, $8-13$, $13-20$ and $20-32$\,kpc respectively.   In all panels, 
a gray scale that varies linearly with the star number is used.
}
\end{figure*}

\begin{figure*}[t]
\begin{center}
\includegraphics[width=4.7cm]{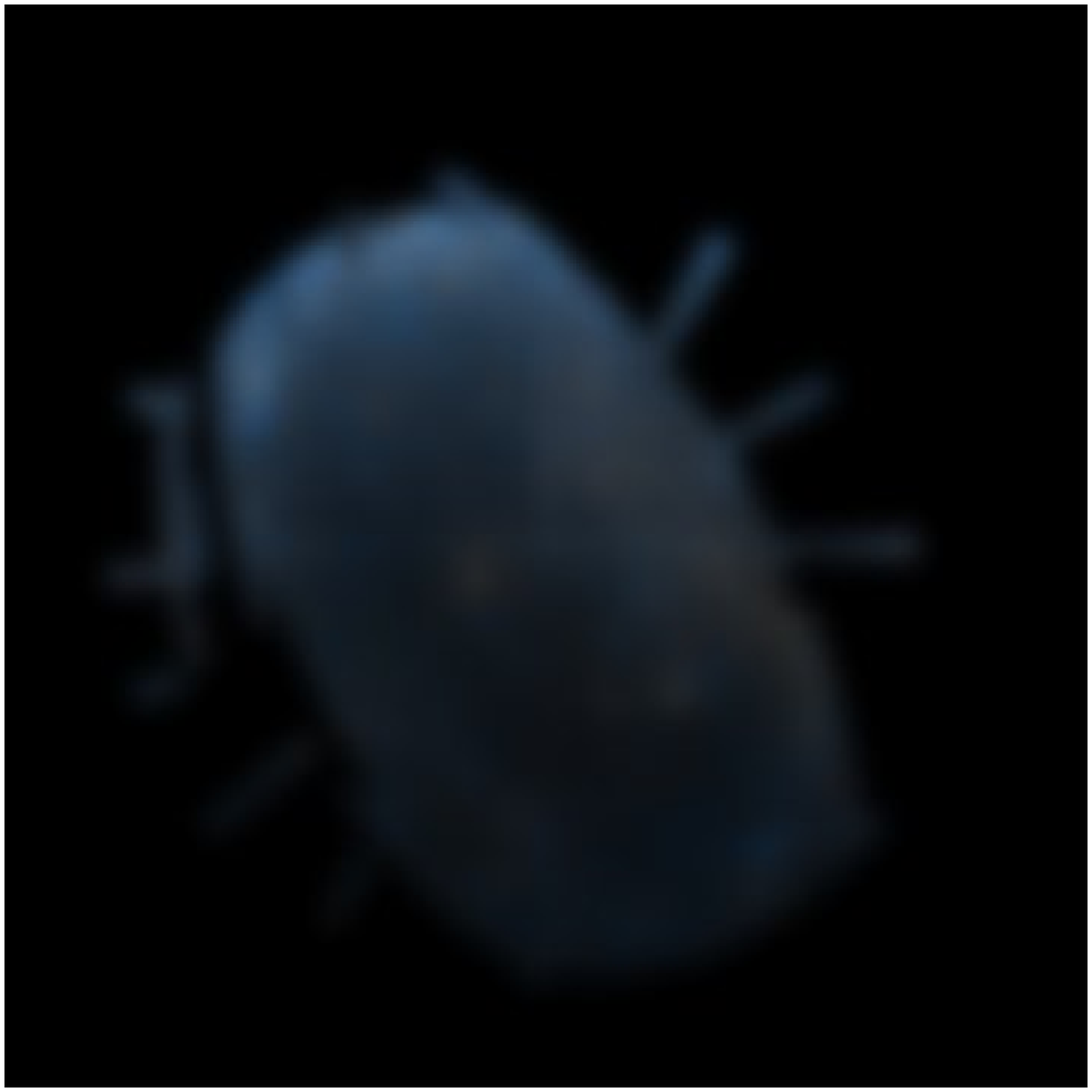}
\includegraphics[width=4.7cm]{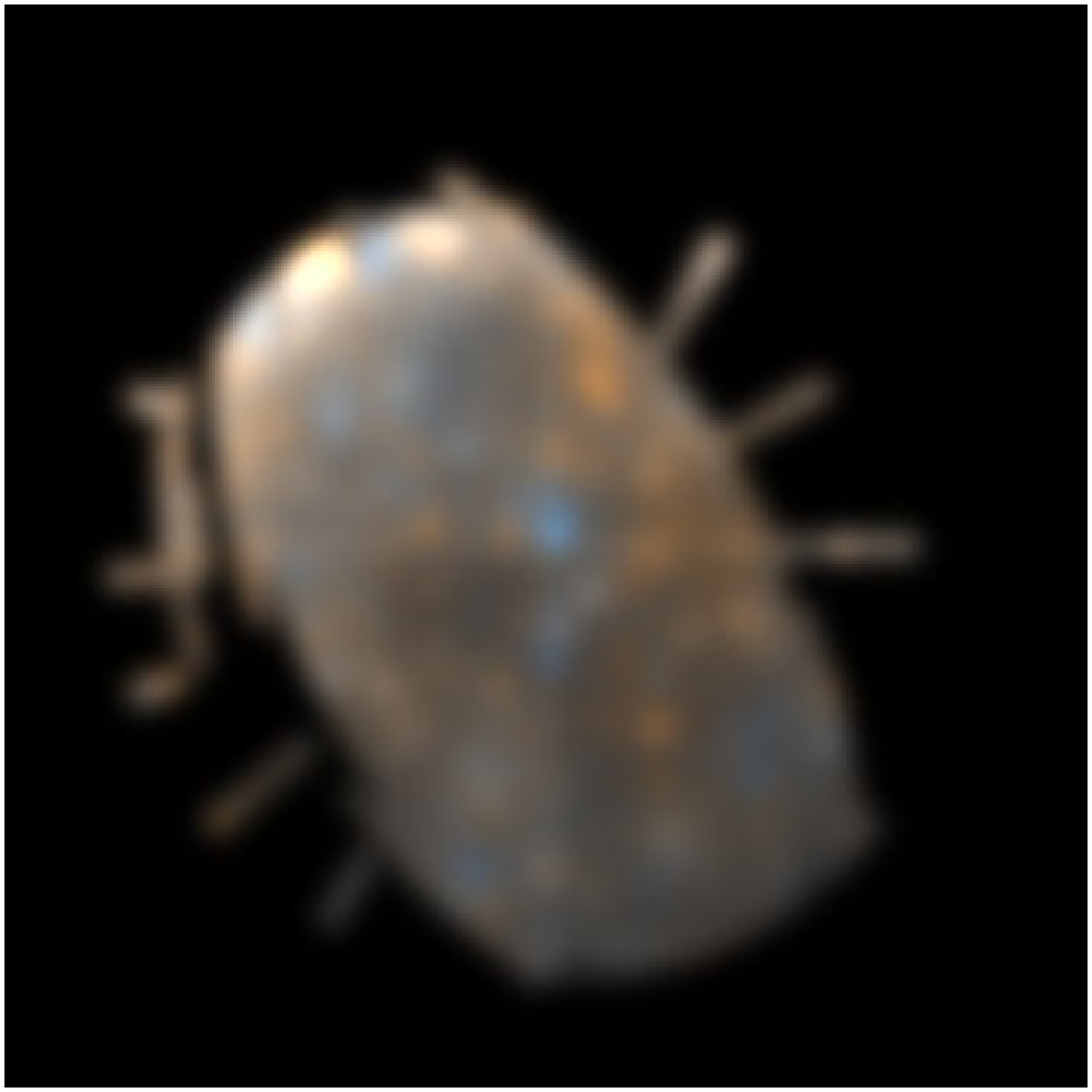}
\includegraphics[width=4.7cm]{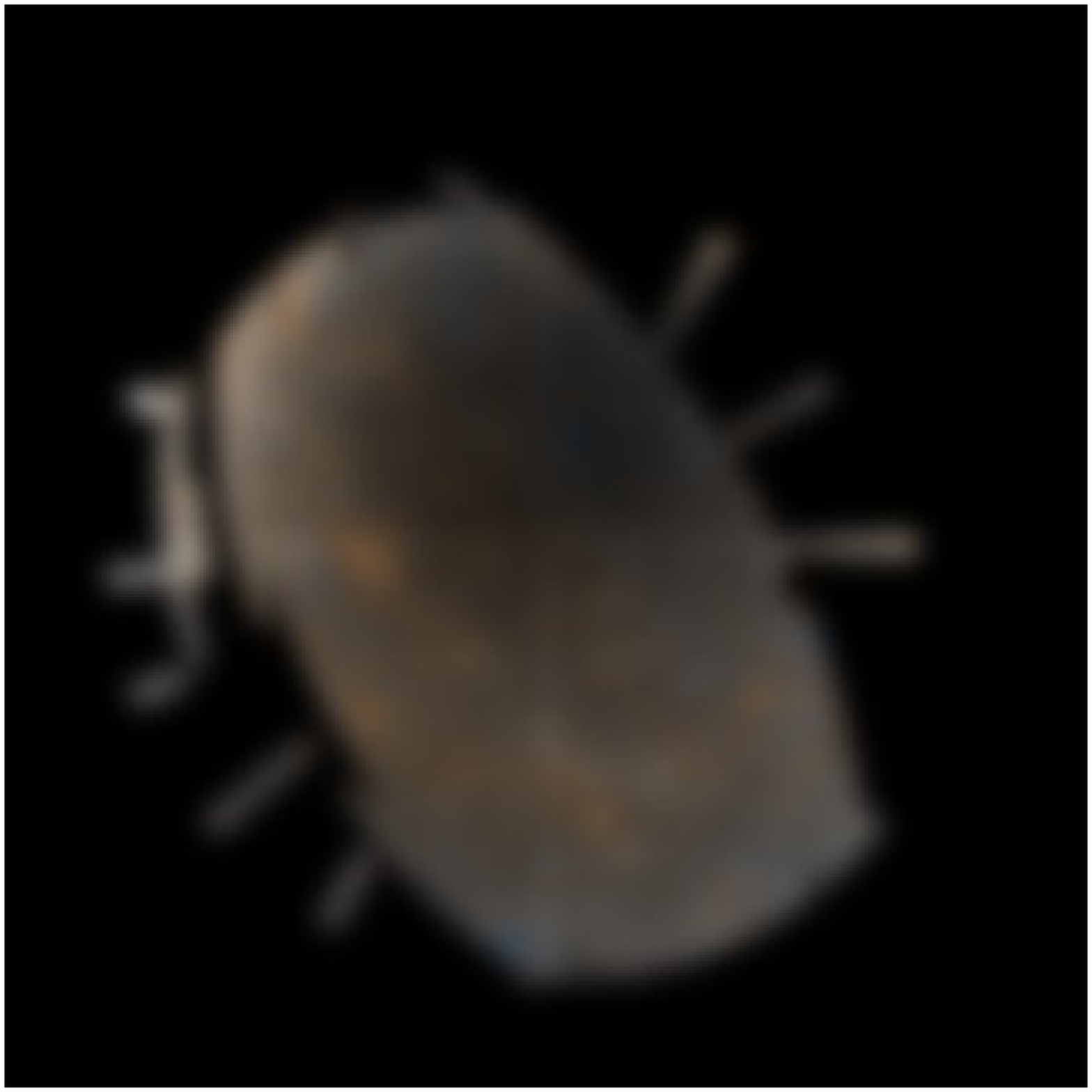}
\includegraphics[width=4.7cm]{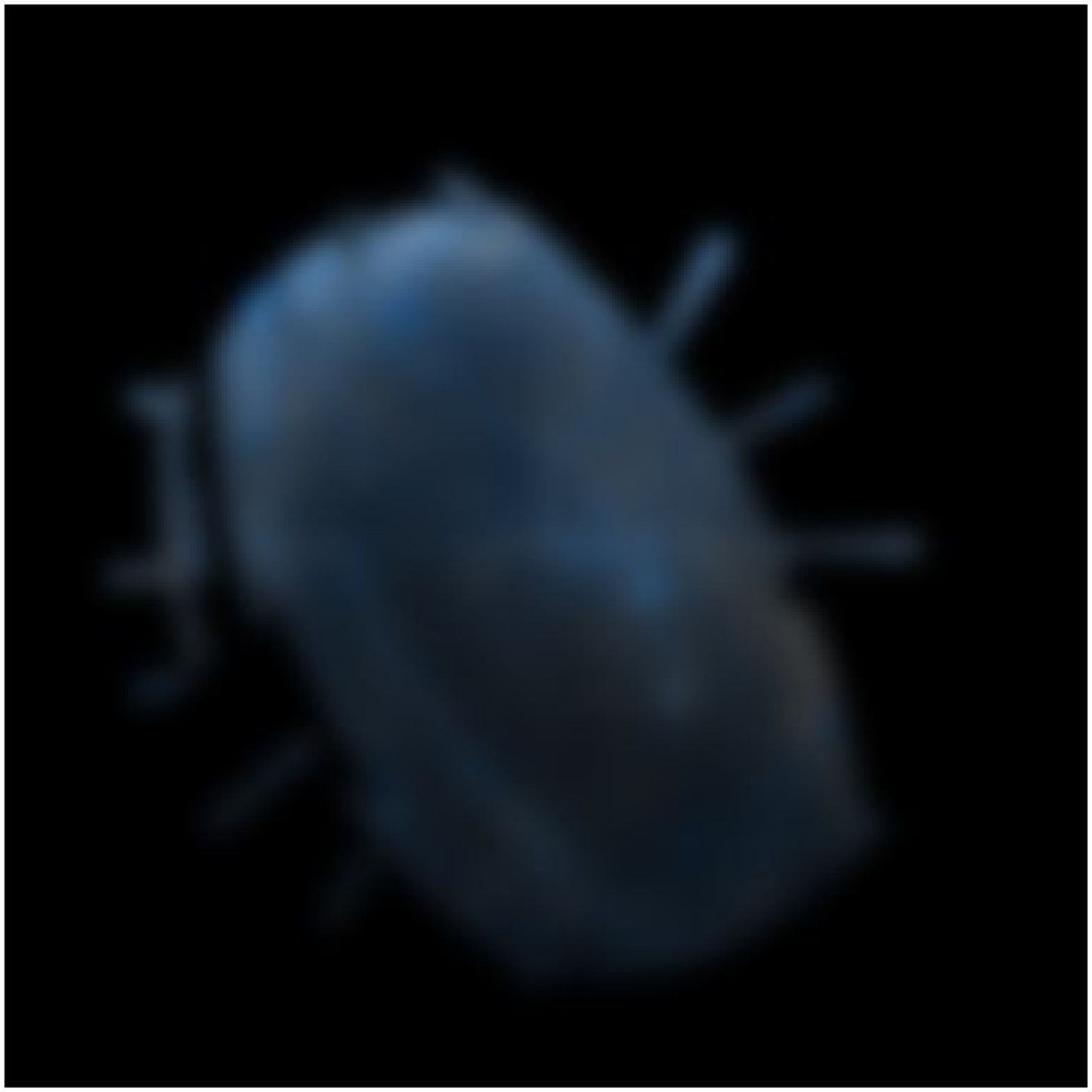}
\includegraphics[width=4.7cm]{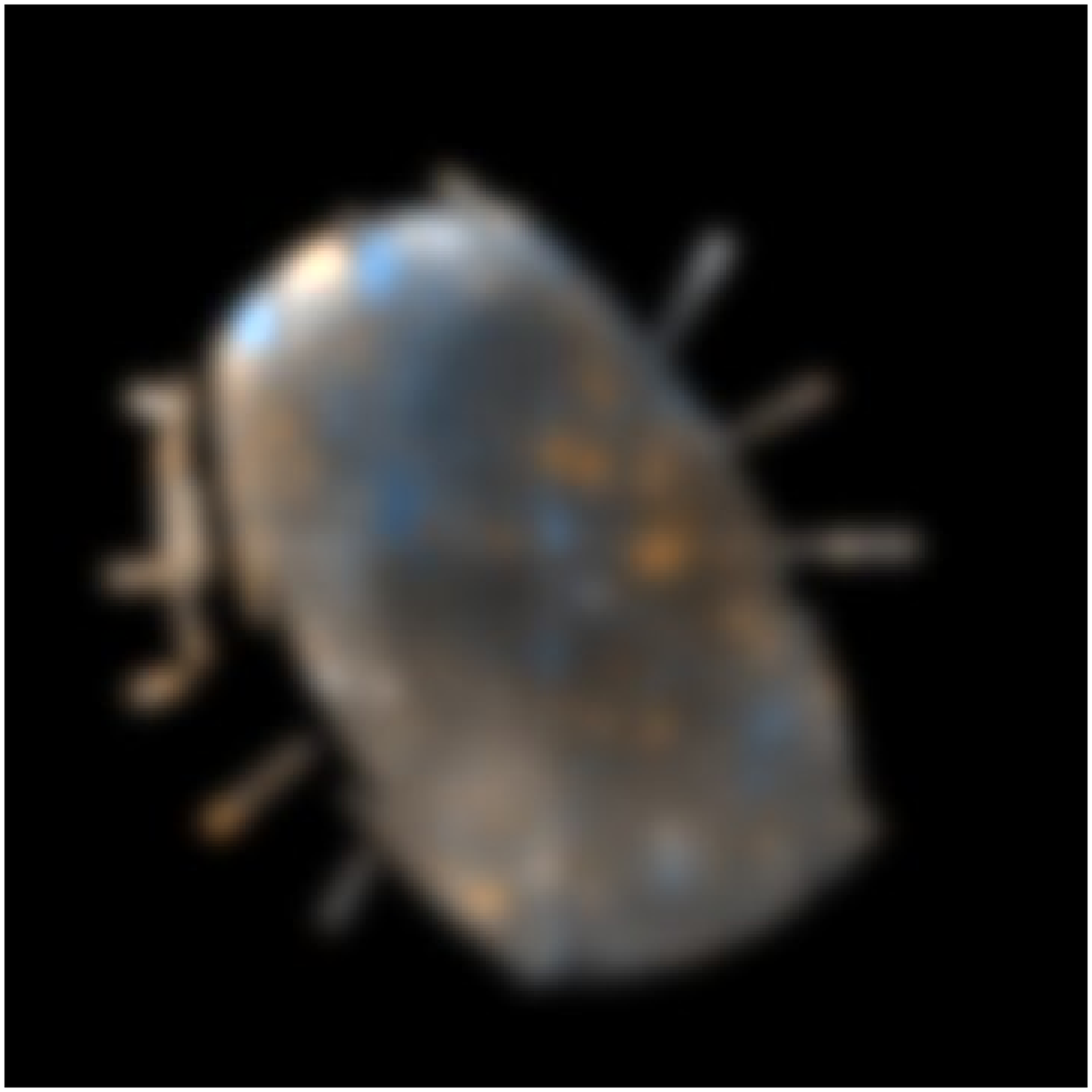}
\includegraphics[width=4.7cm]{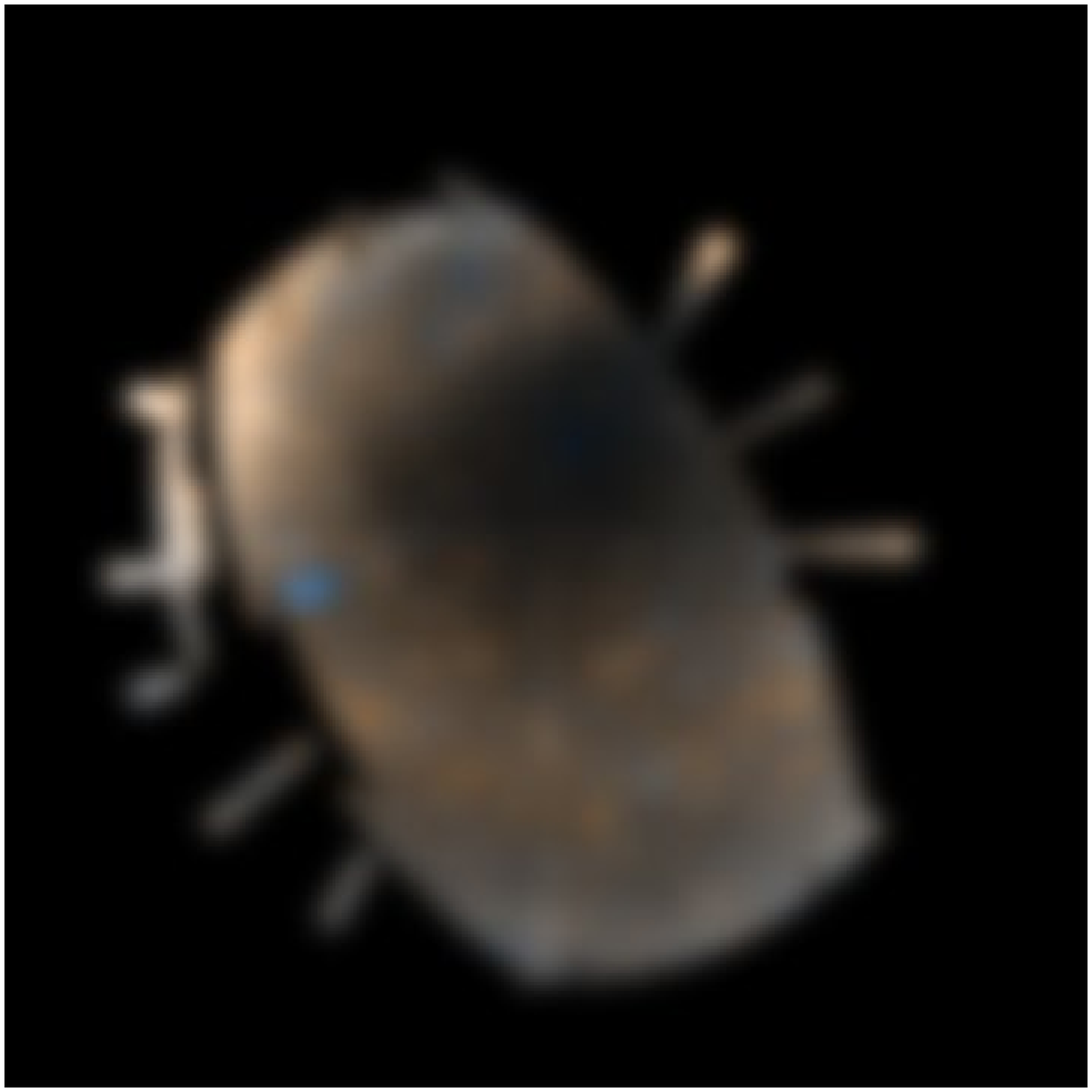}
\includegraphics[width=4.7cm]{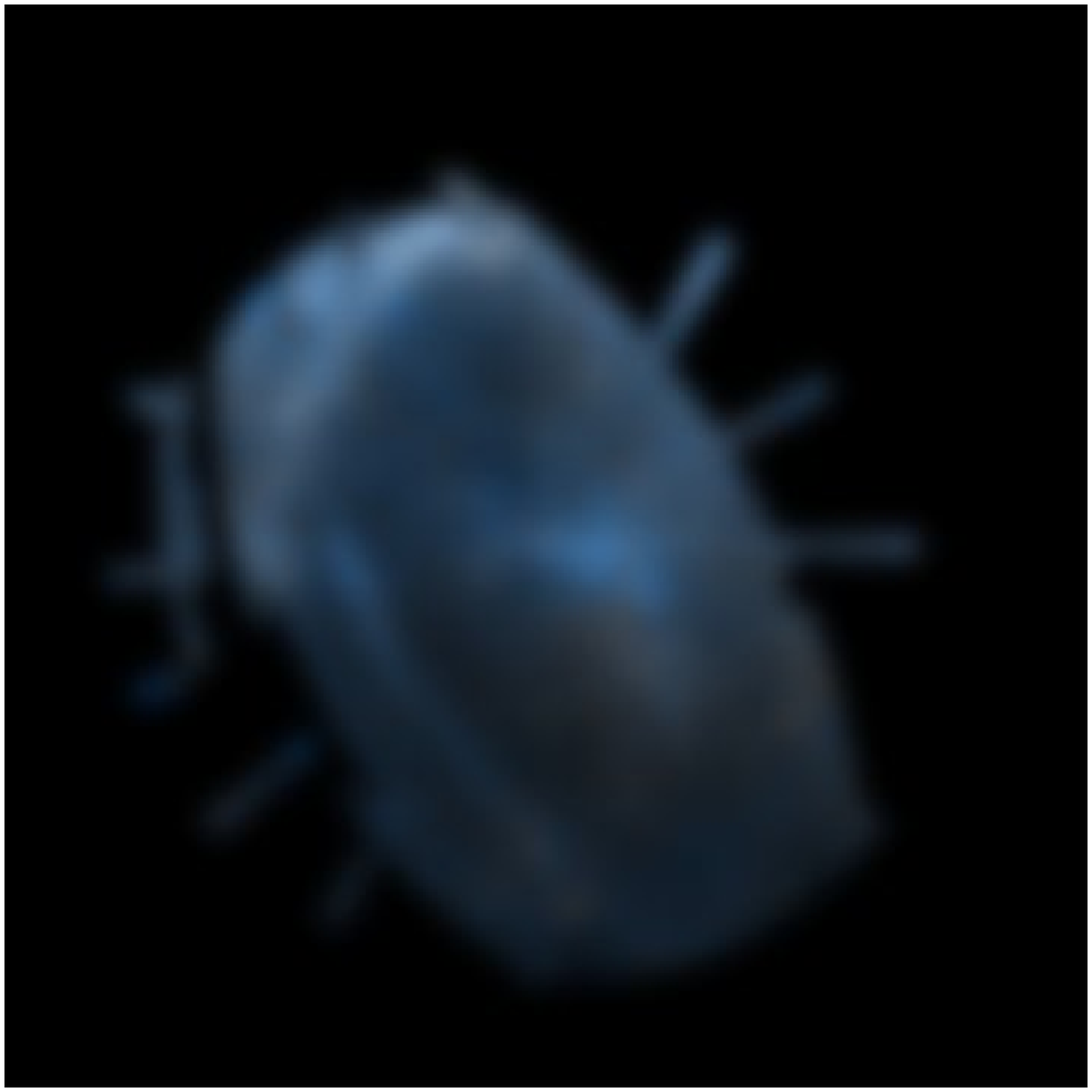}
\includegraphics[width=4.7cm]{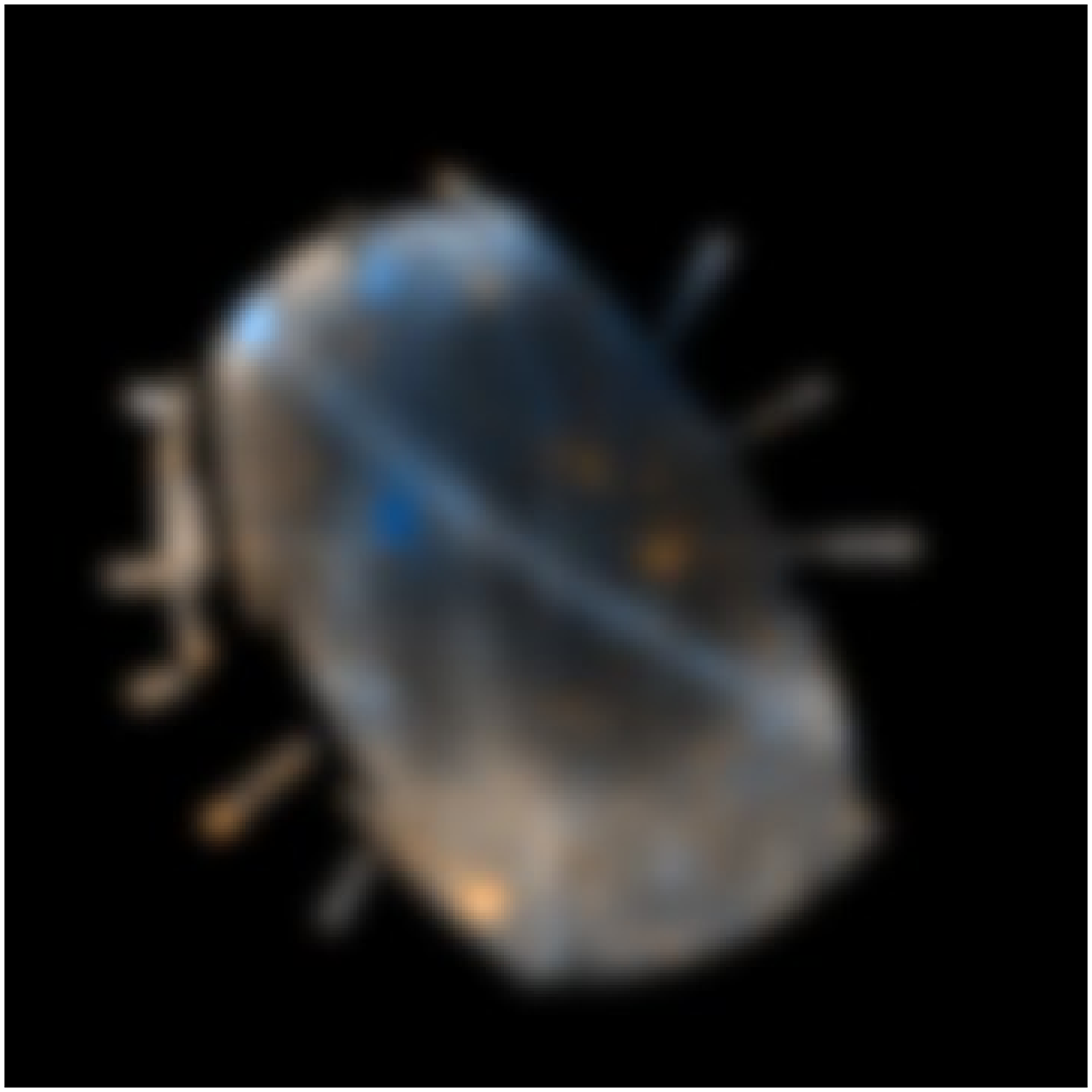}
\includegraphics[width=4.7cm]{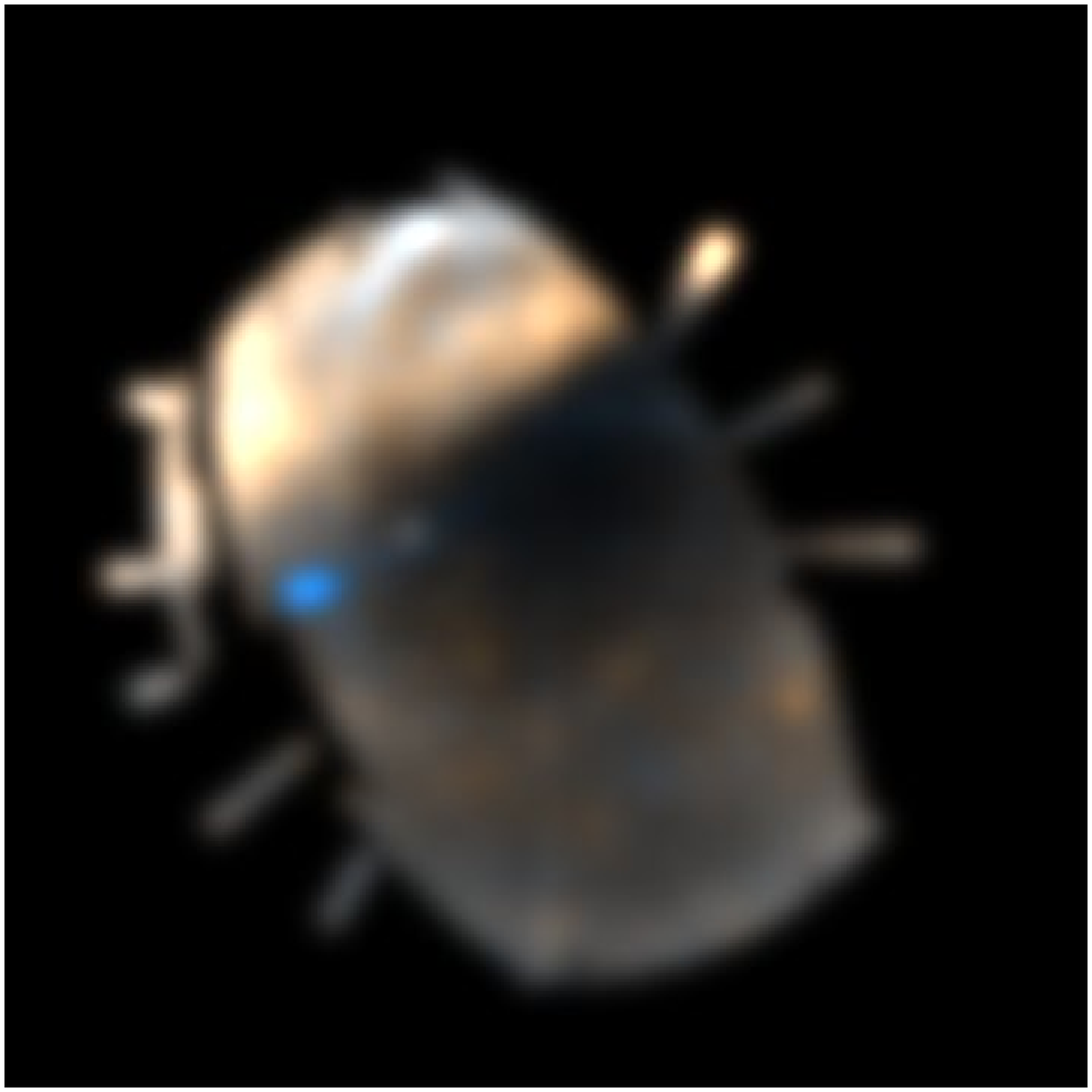}
\includegraphics[width=4.7cm]{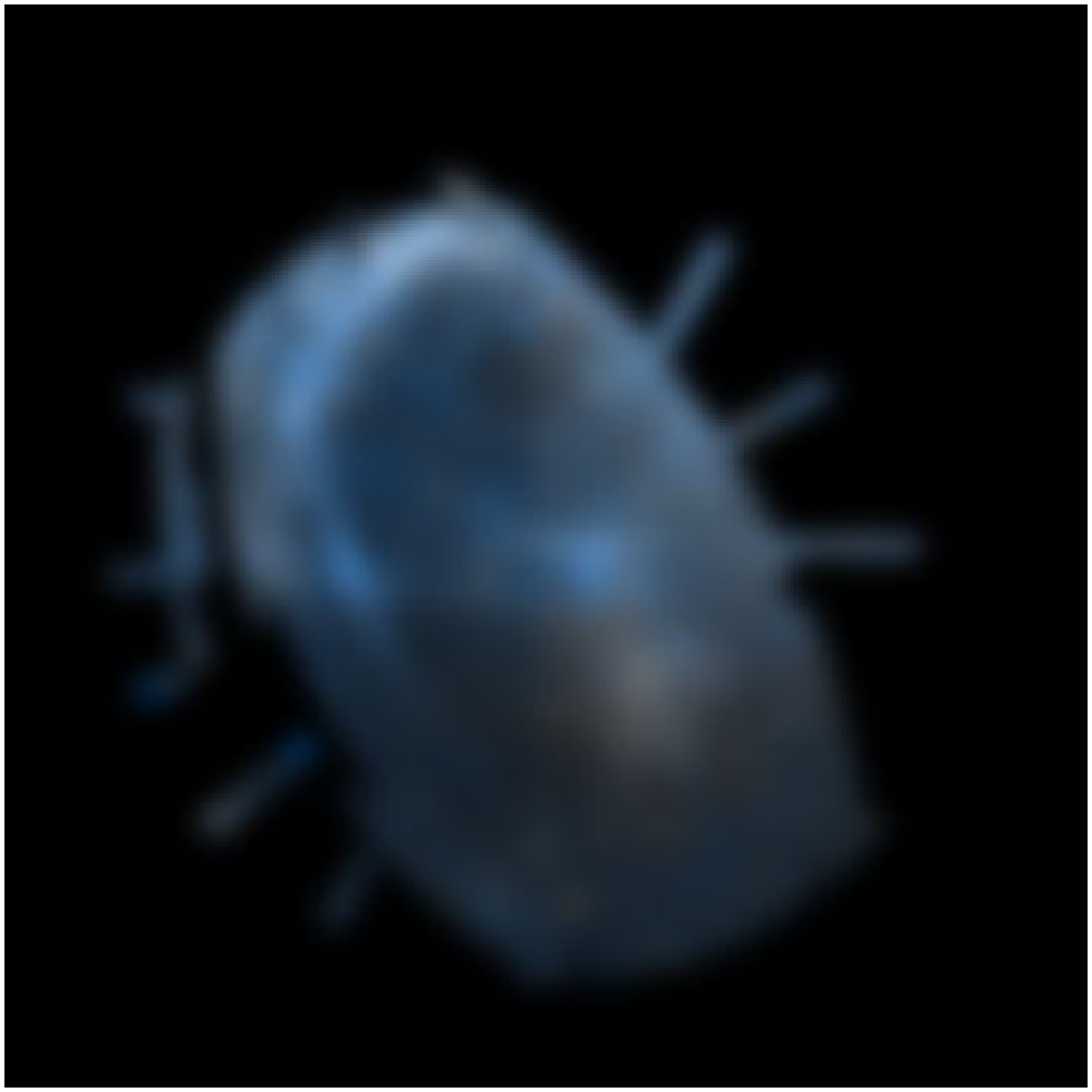}
\includegraphics[width=4.7cm]{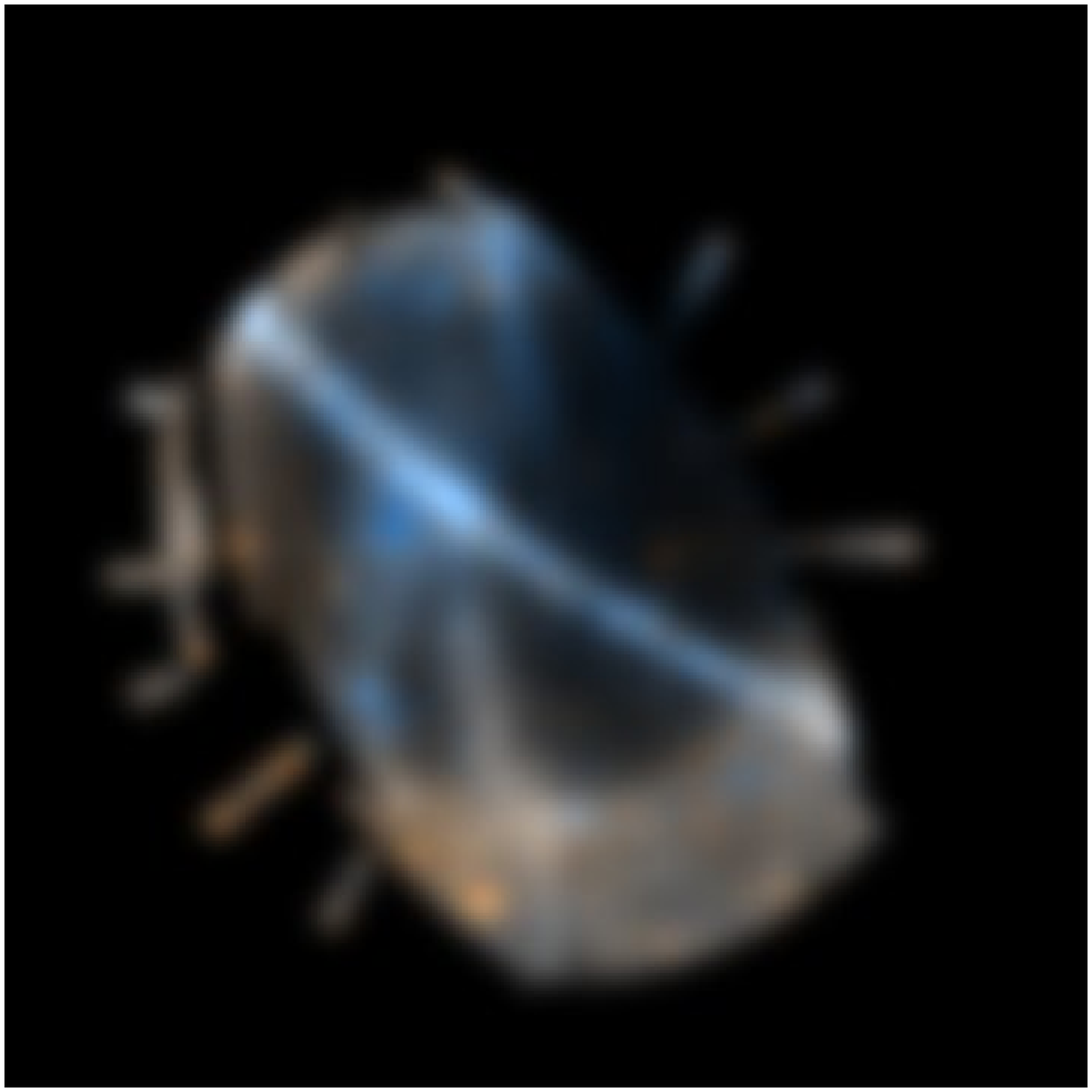}
\includegraphics[width=4.7cm]{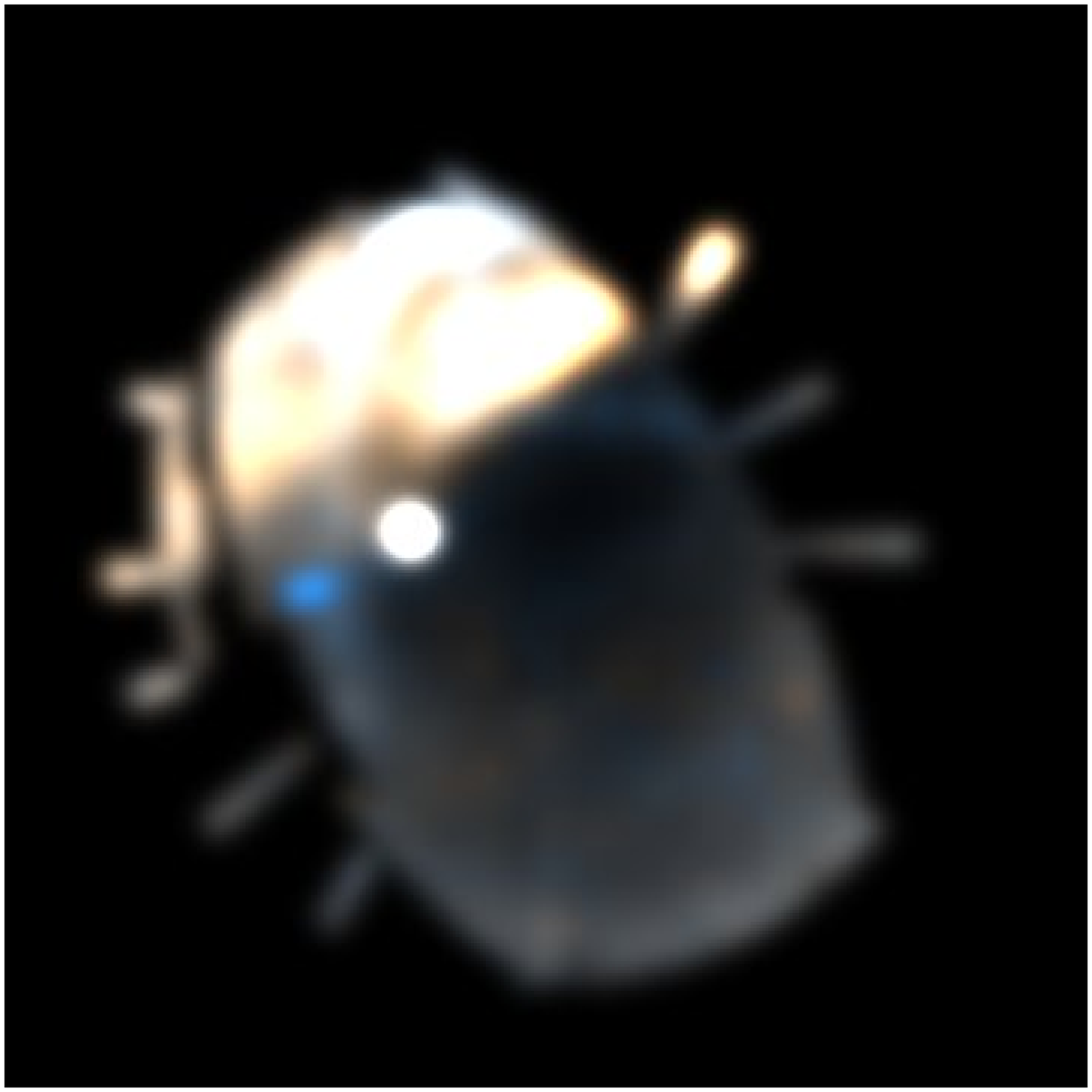}
\end{center}
\caption{\label{fig:theorymap}
A color representation of the relationship between structure and 
metallicity in simulated stellar halos of 
\protect\cite{bullock05}.  The three columns show stellar halos with 
relatively low, average, and large amounts of substructure.  
The 4 rows show the simulated halos convolved with the MSTO 
absolute magnitude distribution and split into the 
same distance bins as used for the 
SDSS data: $13.5 \le m-M < 14.5$, $14.5 \le m-M < 15.5$, 
$15.5 \le m-M < 16.5$, and $16.5 \le m-M < 17.5$, or distances of
$5-8$, $8-13$, $13-20$ and $20-32$\,kpc respectively. 
The intensity scale is linear and scales with the number of stars; the color
scale shows metallicity, where blue denotes ${\rm [Fe/H]} = -1.5$ 
and red denotes ${\rm [Fe/H]} = -0.5$. 
}
\end{figure*}

We use publicly available catalogs
derived from the Sloan Digital Sky Survey's (SDSS) Data Release
Seven \citep[DR7;][]{dr7} to select BHB and MSTO stars 
for further study.
The SDSS is an imaging and spectroscopic survey that has
so far mapped $\sim 1/4$ of the sky. Imaging data are produced simultaneously
in five photometric bands, namely $u$, $g$, $r$, $i$, and
$z$~\citep{fu96,gunn98,hogg01,gunn06}. The data are processed through
pipelines to measure photometric and astrometric properties
\citep{lupton99,st02,smith02,pier03,iv04,tucker06} and to select targets for
spectroscopic follow-up \citep{blanton03tiling,strauss02}.  

To obtain full area coverage and maximal distance range, we must rely
on BHB star samples selected by $ugr$ color information alone.  
One of the well-known practical challenges of using BHB stars for stellar halo 
science is that genuine BHB stars are outnumbered by intrinsically fainter
blue straggler stars (as discussed at length by \citealp{kinman94}, \citealp{wilhelm99}, \citealp{clewley02}, \citealp{sirko04}, \citealp{kinman07}, and \citealp{xue08}; some authors refer to these as a blue plume, e.g., \citealp{bellazzini06sgr}).  Fig.\ \ref{fig:gc} illustrates this issue\footnote{Because NGC 5024 is particularly BHB star rich, the blue stragglers do not dominate in this figure.  When translated to the global halo population, and accounting for the stellar halo density profile and different distances of BHB and blue straggler stars at a given magnitude, BHB stars are typically outnumbered by a significant factor by blue stragglers.}, showing
BHB stars at $g \sim 16.85$ and less luminous blue straggler stars with similar
colors at $18 \la g \la 20$.  These blue straggler stars are a common feature of globular clusters, dwarf galaxies and the stellar halo \citep[e.g.,][]{sandage,bellazzini06sgr,preston00,carney01}.  Blue stragglers are somewhat enigmatic, as they are intermediate mass main sequence stars in (old) populations where the MSTO is at lower masses.  It is thought that there are at least two possible formation routes: stellar collisions in dense star clusters \citep{hills76}, and mass transfer in the evolution of close binaries (\citealp{mccrea64}, or potentially in triple star systems; \citealp{perets09}) in low-density environments (where the latter low stellar density route is more relevant here; see, e.g., \citealp{preston00}, \citealp{mathieu09}, \citealp{knigge09}).  

\subsection{Color selection of BHB stars}

Choosing stars in the broad 
color cut broad cuts $-0.5<g-r<0$ and $0.8<u-g<1.6$ gives a sample 
that is $\ga$3:1 blue stragglers to BHB stars for the full
$g \la 19.5$ sample of \citet{xue08}, and 60\% blue stragglers for
the $g \la 18$ subsample of interest in this paper.  This ratio is 
based entirely on the sample of stars with SDSS spectroscopy as reported
in \citet{xue08}.  The algorithm used to select stars for SDSS spectroscopy
changed a number of times during the survey, and does not select uniformly
in color-color space; these contamination fractions are affected by 
this selection (we present alternative estimates of contamination fractions
derived in other ways later).  Yet, it is clear that such a level of 
contamination is a significant challenge for quantitative analysis
of the BHB star population using photometry alone.

Our goal here is to combine the extensive SDSS spectroscopy with 
SDSS photometry to devise color selection criteria that are significantly 
improved in their trade-off
between BHB purity and sample size (similar in spirit to the 
'stringent' color cut of \citealp{sirko04}; see also 
\citealp{clewley05} and \citealp{kinman07} for selection of BHB stars 
from $ugr$ photometry).   
Using a calibration sample of BHB star candidates with spectra
(\citealp{xue08}, following the methodology of \citealp{sirko04}),
we have determined the probability that a BHB candidate
with given $u-g$ and $g-r$ color is a spectroscopic BHB star; 
the rest of the candidates
are other blue stars, and are presumably mostly blue stragglers.
Fig.\ \ref{fig:selection} illustrates this process, showing 
stars with $g < 18$ (appropriate for BHB stars in the Heliocentric distance range 
considered here) and with spectra from the SDSS \citep{xue08}.  
Blue data points 
show stars that are very likely to be BHB stars on the basis of their 
spectra (a contamination of much less than 10\% has been 
argued by \citealp{xue08} and \citealp{sirko04} at these $g<18$ limits).
The thick contour outlines the region of 
color-color space where the fraction of BHB candidates that 
are spectroscopically-classified BHB stars is $>50$\% 
and there were more than 16 stars in a bin of 
0.025$\times$0.04 mag (to ensure a high enough number of stars to measure the BHB star fraction with some fidelity).  
This region is well-approximated 
by the selection region shown in red in Fig.\ \ref{fig:selection}: 
$0.98<u-g<1.28$, $-0.2<g-r<-0.06$ and 
excluding the region with $([u-g-0.98]/0.215)^2 + ([g-r+0.06]/0.17)^2 < 1$
(formulated similarly to the 'stringent' color cut of \citealp{sirko04}, 
shown in green).  

\subsection{Completeness and contamination of color-selected samples}

We first quantify the likely completeness and contamination of the color-selected $g<18$ sample that is the main focus of this paper.  We estimate the completeness and contamination in two ways: using the spectroscopic sample, and independent spot checking on BHB star-rich globular clusters.

Our primary estimate of contamination and completeness comes from the 
spectroscopic training sample used to determine the color cuts.
Given the average spectroscopic completeness 
within the $>50$\% contour, and weighted by the number of 
stars in our photometric catalog per bin in color-color space (this to 
a certain extent corrects for the non-uniform color selection of 
stars with SDSS spectra), 
we expect $\sim 67-74\%$ of the BHB candidates within 
the selection region to be 
real BHB stars (where the two ends of the range assume that 
90\% to 100\% of the spectroscopic BHB stars are actually BHB stars, 
following \citealp{sirko04} and \citealp{xue08}).  
Furthermore, we find that the color cuts select $\sim 60$\% of 
the BHB stars in the spectroscopic sample.  

Secondary estimates of contamination and completeness are possible 
through the investigation of BHB stars in globular clusters 
and dwarf galaxies.  
This method has the advantage that an independent indicator of the nature of 
the stars is available through their position on the CMD, but have the disadvantage that there are very few clusters in the relevant distance range, and that the number of BHB stars per cluster is typically very modest (dwarf galaxies were also tested, but are more distant, and give results qualitatively consistent with the picture presented here but are of limited quantitative use).  
In Fig.\ \ref{fig:gc}, we illustrate this method 
to the BHB star-rich globular cluster NGC 5024.
Stars selected
as BHB candidates by the broad cuts $-0.5<g-r<0$ and $0.8<u-g<1.6$
are shown as crosses, while those with $>50\%$ probability 
as estimated by our method are shown as diamonds.   
There are 30 actual BHB stars in this sample at the distance of NGC 5024
(defined as having $-0.4<g-r<0$ and $16.6<g<17.1$); 
our method recovers
11 of them with only 2 contaminants (i.e., a completeness of
$\sim 37\%$ and a contamination of $\sim 15$\%, consistent with 
our expectation of $\la 30$\% contamination).  There are indications that 
NGC 5024 may have an unusually low fraction of recovered stars; NGC 5053
has 5/7 ($\sim 70$\%) BHB stars recovered (with no contaminants).
  
Considering both tests together, we conclude 
that this procedure, when applied to $g<18$ stars with SDSS-quality photometry,
gives a sample of $\sim 50$\% of 
the real BHB star population with $\la 30$\% contamination, a considerable 
improvement over the $\sim$60\% contamination given by much more simplistic 
photometric cuts.

While in this paper we restrict our attention to BHB stars with $g \la 18$, 
we characterize the likely performance of these cuts for fainter samples
with colors characteristic of BHB stars.  To estimate the run of photometric 
uncertainty as a function of measured apparent magnitude in $u$, 
$g$ and $r$-bands, we used repeat imaging of the 
SDSS stripe 82 as tabulated by \citet{bramich}.
We then imposed the typical uncertainties of stars 
with BHB colors at $g \sim 19$ and $g\sim 20$ on the sample 
of real $g<18$ stars used to derive the BHB star selection, and 
checked how the fraction of the real BHB stars that are 
recovered, and the contamination by BS stars, 
should depend on apparent magnitude.  Recall for the $g<18$ stars, 
the recovery/contamination fractions for the spectroscopic sample 
are $(60\%,25\%)$, i.e., the cut isolates
60\% of the population with only 25\% contamination.  For $g = 19$ and $g=20$, 
our estimates of the recovery/contamination 
fractions are $(40\%,40\%)$ and $(20\%,50\%)$ 
respectively, i.e., at $g=19$ the situation is  
improved over the 60\% contamination in the 
broad $-0.5<g-r<0$, $0.8<u-g<1.6$ color bin by applying the
cuts described here, but at $g=20$ the cuts throw out most of 
the genuine BHB stars without improving contamination significantly. 
These fractions are consistent with our investigations of 
the performance of cuts on more distant BHB-rich targets, e.g., 
the Sextans or Draco dwarf galaxies (1/4 of the BHB stars recovered, 
at $g \sim 20$).  

\subsection{Estimates of BHB star absolute magnitude}

We estimate the absolute magnitudes for the BHB star candidates
by interpolating the $g-r$-dependent $M_g$ calibration from
Table 2 of \citet{sirko04} with [Fe/H]$=-1$ 
(the [Fe/H]$=-2$ \citealp{sirko04} calibration is different by $<0.03$\,mag).
Tests of the quality of this calibration are encouraging; 
the rms of the color-derived absolute magnitudes of BHB stars in 
globular clusters is $\sim 0.1$\,mag 
(a measure of the absolute magnitude scatter at a single metallicity), 
and the overall distance modulus 
estimates of the clusters and dwarf galaxies agree
with those determined by independent means to $\la 0.1$\,mag 
(a rough constraint on the degree of metallicity dependence 
of the BHB absolute magnitude calibration; these results are 
reported in full in \citealp{ruhland10}).  

We also compared the calibration of BHB absolute magnitude 
from \citet{sirko04} with the metallicity-dependent BHB star models from 
\citet{dotter07} and \citet{dotter08}.
These models make use of more realistic input physics than the modeling 
from \citet{sirko04}; however, the $ugr$ colors of 
BHB stars from \citet{dotter08} do not match the colors of SDSS BHB stars
particularly well (they are offset in particular in $u-g$ color 
by $\sim 0.1$\,mag, 
and may be offset by a smaller amount in $g-r$ color; in contrast, 
the more rudimentary modeling by \citealp{sirko04} was designed to match 
the colors well).  These offsets in color prohibit a quantitative
assessment of which stars would fall in the selection region; selecting
instead those BHB stars reddest in $u-g$ at a given $g-r$ 
(a qualitative version of our selection method), corresponding to 
metallicities of ${\rm [Fe/H] \la -1.5}$, gives a full range in 
$\Delta M_g \la 0.3$\,mag 
at a given $g-r$.
Taken together, the tests of absolute magnitude calibration on
clusters and dwarf galaxies in conjunction with 
the \citet{dotter08} metallicity-dependent BHB star models paints a picture
in which the absolute magnitudes adopted for our BHB candidates are 
accurate to $\la 0.1$\, in overall zero point, and will 
scatter $\la 0.2$\,mag at worst -- far less than the $\sim 0.9$\,mag
absolute magnitude scatter of our MSTO sample.

The final sample 
of high-probability BHB stars selected using the 50\% color contour
in Fig.\ \ref{fig:selection} and with $g < 21$ is 
more than 14000 stars in the North Galactic Cap selection region we use in this paper ($b>20$ for $180<l<240$, and $b>30$ for $l\le 180$ and $l\ge240$, dictated largely by the SDSS coverage), and more than 9000 stars 
in the Heliocentric distance range
5--30\,kpc (corresponding approximately to $g \la 18$).

\subsection{Selection of MSTO stars, and comparison of MSTO and BHB stars}

Following \citet{bell08}, we select over 4.3 million MSTO stars in the
foreground extinction-corrected color range
$0.2<g-r<0.4$ and $18<r<22$; this color and magnitude 
range was selected empirically to 
encompass the most densely-populated bins of color space for 
the halo MSTO stars (see Fig.\ 1 of \citealp{bell08}).  Examination 
of globular clusters and model CMDs 
demonstrates that MSTO stars selected in this fashion have 
a roughly Gaussian distribution of absolute magnitudes with a 
mean $M_r \sim 4.5$ and a rms $\sigma_{mag,MSTO} = 0.9$
(see Fig.\ 2 of \citealp{bell08}).  
By choosing a particular color range for MSTO star selection, 
we are in principle susceptible to changes in the main-sequence 
stellar population from location to location in the stellar halo
(through variation in the metallicity and age).  In practice, 
this turns out to be a minor concern, as the color of the stellar halo MSTO 
(the blue edge of the turn-off in particular) varies
little across the SDSS area; the stellar population 
in the stellar halo varies little enough that it is essentially
undetectable using MSTO color alone. 

Fig.\ \ref{fig:sgr} illustrates the properties of the stellar 
sample in a particular line of sight towards the Sagittarius tidal 
tail.
This shows the CMD of (primarily halo) 
stars in a 4$\arcdeg$ radius 
cone around (RA,Dec)$=$(204,5).  MSTO stars in the halo are very numerous, 
having $0.2<g-r<0.4$ and $g>18$ (brighter MSTO stars with $g<18$ have $g-r \sim 0.4$, representing more metal-rich thick disk stars).  
BHB candidates are selected by virtue of their $ugr$ colors
using the method illustrated in Fig.\ \ref{fig:selection}, and are shown 
as crosses.  Because BHB stars have a small scatter in absolute magnitude, 
BHB star-rich halo overdensities (at $m-M \sim 16.5$ and $m-M \sim 18.3$)
are clearly-defined.  The corresponding MSTO overdensities, about 
4 magnitudes fainter, are not anywhere near as clearly
defined owing to the larger spread in MSTO absolute magnitudes. 
In this paper, we will be analyzing the two populations in concert
as an easily-measurable but largely qualitative diagnostic for 
stellar population variations in the stellar halo of the Milky Way.

\section{Population Variations in the Stellar Halo of the Milky Way} \label{sec:bhbmsto}

Analyzing the ratio of BHB to MSTO stars requires that 
we compare the stars in the same `effective volume'.
We convolve the BHB star distance moduli with a 
Gaussian with width $\sigma_{mag,MSTO} = 0.9$\,mag 
(referred to hereafter as the degraded BHB sample) to yield
the same scatter between degraded distance modulus $(m-M)_{degraded}$ and true
$(m-M)_{true}$ as is estimated for the MSTO stars.  
Then we can compare the `distance smoothed' distribution of BHB 
stars, with $(m-M)_{degraded}$, directly and 
quantitatively to the distribution of MSTO stars with 
the same $(m-M)_{MSTO}$ range (i.e., BHB stars are compared with 
MSTO stars that are $\sim$4\,mag fainter)\footnote{It should be noted
that this shift in magnitude is important also when considering 
blue straggler contamination.  Blue stragglers are 1-3 mag fainter
than BHB stars, therefore any blue straggler star contamination should show up
at larger equivalent distance modulus (because the blue stragglers 
are being treated
as brighter BHB stars).  }.
The results are illustrated in 
Fig.\ \ref{fig:map}. 

Fig.\ \ref{fig:map} displays various aspects of the comparison 
between BHB and MSTO star distributions, and we work through 
them in turn.  The center column of panels
shows the distribution of MSTO stars in four distance modulus
slices: $13.5 \le m-M < 14.5$, $14.5 \le m-M < 15.5$, 
$15.5 \le m-M < 16.5$, and $16.5 \le m-M < 17.5$, or distances of
$5-8$, $8-13$, $13-20$ and $20-32$\,kpc respectively, smoothed 
in angular position with a 
Gaussian kernel with $\sigma = 0.5\arcdeg$.  The gray scale shows
the number of stars in $0.5\arcdeg \times 0.5\arcdeg$ bins 
with a linear mapping 
between intensity and stellar density.  Black areas lack SDSS DR7 coverage.
These panels show various previously described features of the 
Milky Way stellar halo
\citep[e.g.,][]{fos,newberg07,bell08}: superimposed on a 
relatively smooth distribution of stars is the Sagittarius 
stream \citep[e.g.,][]{ibata95,majewski03,fos} stretching 
from the Galactic anticenter ($m-M \sim 16$) to the
Galactic center ($m-M \sim 17$).  Near the North Galactic Pole (NGP) at
$m-M \la 16$ is the diffuse Virgo overdensity 
\citep[e.g.,][]{duffau06,newberg07,juric08}.
At $m-M \la 15$ in the Galactic anticenter 
direction at low galactic latitudes is the 
low-latitude structure 
\citep[also referred to as the Monoceros stream; e.g.,][]{newberg02,pena05,momany06}.
At $l \sim 45\arcdeg$ and $b \sim 45\arcdeg$ is the diffuse Hercules-Aquila
structure \citep{her_aq}.  One can also see a variety of less prominent
features: globular clusters as `hot pixels', and at $m-M \sim 17$ 
stretching from $(l,b) \sim (250,50)$ to $(l,b) \sim (160,40)$ 
is the Orphan Stream \citep{orphan,grillmair_orphan}.

In the left-hand column of panels in Fig.\ \ref{fig:map}, we show the 
BHB star distribution in the same distance bins, and convolved with 
a distance modulus kernel of width $\sigma_{\rm MSTO} = 0.9$\,mag to 
match the MSTO absolute magnitude scatter; an angular smoothing 
of $\sigma = 1\arcdeg$ is applied to suppress the much larger shot noise 
due to the 100-fold smaller sample.  
Thus, the left-hand panels reflect BHB star maps of the 
{\it same halo distance slices} as the MSTO panels.  
One can see some similarities, as well as some striking differences, 
between the distributions of the BHB and MSTO stars.  These
differences are quantified and illustrated in the right-hand panels, 
which show a color representation of the
BHB to MSTO star ratio.  
Here, the distributions of BHB and MSTO stars are 
smoothed by a Gaussian kernel with an angle of 
$\sigma = 6\arcdeg$ to increase S/N.  
A bluish-white color denotes BHB/MSTO ratio of $>1/50$ and 
red denotes BHB/MSTO $<1/100$.  The intensity 
scales with the number of stars (dark shades show 
regions with few stars, whereas
light shades denote regions with many stars).  
Poisson noise is significant on small scales, 
but contributes less than 3\% on the scale of 
large structures such as the low-latitude stream.  The 
primary uncertainties are systematic in origin: changing 
the probability threshold for a BHB star to be included 
in the sample from 50\% to 30\% tests the influence of BHB candidate purity and blue straggler contamination on the results.  
Such a change produces differences of $\sim 10$\%  in relative
BHB/MSTO ratio (i.e., variations in the BHB/MSTO from place to 
place on the map)\footnote{Of course, the absolute BHB/MSTO ratio changes
by 30\%, as the number of BHB star candidates included in the analysis
is increased by reducing the BHB probability threshold; such a change
in absolute ratio is not important for our purposes.}.  Another source 
of systematic error is possible variations in MSTO absolute magnitude
or scatter
as a function of position.  Fig.\ 2 of \citet{bell08} shows
that there are some variations in MSTO absolute magnitude
and rms from globular cluster to globular cluster, driven by differences
in stellar population age/metallicity.  Such variations
may be present in the stellar halo.  We have tested the influence of 
potential variations in MSTO absolute magnitude
(a $\la 15\%$ effect in BHB/MSTO ratio for variations in MSTO absolute 
magnitude of 0.5 mag or less) and scatter in the luminosity of 
MSTO stars (a few percent effect in BHB/MSTO ratio for 20\% changes in the 
assumed MSTO scatter).

The right set of panels in Fig.\ \ref{fig:map} demonstrate clearly 
the key result of this paper: there are significant variations
in the BHB/MSTO ratios in the stellar halo of the Milky Way.
Perhaps more importantly, these variations coincide spatially with 
different known halo structures, as 
expected if the various overdensities were composed of the 
debris from the disruption of individual progenitors.
Some prominent structures (e.g., the low-latitude structure) appear to be
almost completely devoid of BHB stars (as illustrated by 
the red band at low galactic latitudes in the Galactic anticenter
direction in the upper right-hand panel of Fig.\ \ref{fig:map}; 
BHB/MSTO $\sim 1/90$).  
Other structures, e.g., the Virgo overdensity at 
$l \sim 300\arcdeg$ and $b \sim 70\arcdeg$ or the Hercules-Aquila
overdensity at $l \sim 45\arcdeg$ and $b \sim 45\arcdeg$, appear to be 
considerably richer in BHB stars (both structures have BHB/MSTO $\sim 1/40$). 
Such differences are highly significant; recalling that the primary
uncertainties are systematic, we find that Virgo and Hercules-Aquila are
a factor of 2 to 2.5 richer in BHB stars than the low-latitude structure (where the range includes the effects of the random and systematic uncertainties discussed above).
Strikingly, parts of the Sagittarius
tidal stream (near the NGP and at $m-M \sim 17$) appear
to be relatively rich in BHB stars (BHB/MSTO $\sim 1/55$), while the part at $m-M \sim 16$ and 
$l \sim 200\arcdeg$ and $b \sim 50\arcdeg$ is considerably poorer
in BHB stars (BHB/MSTO $\sim 1/80$), 
implying a population gradient along the stream. 

Fig.\ \ref{fig:map} also shows
that contamination of the BHB sample by blue stragglers 
would not affect our key result.  
Areas with already low BHB/MSTO ratios (the low-latitude stream 
and parts of Sagittarius) would imply even lower
BHB/MSTO in the presence of blue straggler contamination. Furthermore, 
even for structures such as Virgo or Hercules-Aquila,
the association of BHB stars to particular structures already present 
in MSTO stars (and their having characteristic BHB/MSTO ratios) 
implies that blue straggler contamination has a small effect, 
at least for $m-M \la 17.5$.  The only possible exception to this
is the BHB star-rich section of Sagittarius, which lies in a similar
area of the sky and 1-2 magnitudes more distant than the BHB star-rich 
Virgo overdensity.  It is possible that {\it some} of these BHB stars are 
blue stragglers from Virgo, but there are two arguments that blue straggler
contamination is not dominant.
Firstly, the morphology of the BHB stars
in the $\sim 27$\,kpc slice in Fig.\ \ref{fig:map} argues strongly that most
of the BHB star candidates are in fact BHB stars associated with 
Sagittarius --- the distribution of BHB star candidates is elongated
like the Sagittarius tidal stream, and extends past $l=270$ where
there are no stars in the Virgo overdensity (as can be seen in 
the $\sim11$\,kpc slice).  Secondly, there 
are numerous blue stars along those lines of sight that are 
1-2 mag fainter than the BHB stars we identify here; we interpret
these numerous fainter blue stars as the blue straggler content
of Virgo and Sagittarius (this is illustrated 
in \citealp{niederste} and \citealp{ruhland10}). 

\section{Comparison with models of stellar halo formation in a cosmological context}
\label{sec:mod}

One motivation for this analysis was to test further
the hypothesis that the stellar halo is composed of the 
debris from the disruption of dwarf galaxies.  
Cosmologically-motivated simulations of stellar halo formation such 
as those of \citet{bullock05} or \citet{cooper10} 
allow reasonably direct comparison
of the observations and simulations.  
\citet{bullock05} have 
publicly released\footnote{{\tt http://www.astro.columbia.edu/$\sim$kvj/halos/}}
the results of their simulations, including 
the luminosities, ages, metallicities, and element abundances of 
star particles \citep[see also][]{robertson05,font06} in eleven
simulated stellar halos.
Our main observational diagnostic, the spatially-varying 
ratio of MSTO to BHB stars, cannot be predicted directly by the models:
while old
and metal poor populations contain a significant population of BHB
stars, it is not completely
clear what the exact conditions are under which stellar populations are
rich in BHB stars.  An illustration of this challenge
is the second-parameter problem: globular clusters with very
similar metallicities can have very different horizontal branch 
morphologies, where it is unclear what parameter or parameters drive
these variations (e.g., age or Helium content; 
see, e.g., \citealp{catelan05} for a discussion of this issue).  
Obviously, this uncertainty makes a correct BHB prescription impossible 
to implement. 

Bearing in mind the practical difficulty of predicting 
the ratio between BHB and MSTO stars,
we probe instead a metric that is better-defined in the models that gives some 
{\it qualitative} idea of population variations in stellar halos:
stellar metallicity.
We present in Fig.\ \ref{fig:theorymap} a map of 
$r$-band luminosity-weighted stellar metallicity
of three simulated stellar halos (Model numbers 4, 2, and 1 of Fig.\ 13
of \citealp{bell08}, corresponding to halo numbers 8, 5 and 2 respectively from \citealp{bullock05}) in the same distance bins as the observations
(and convolved with 
the same absolute magnitude scatter of 0.9\,mag as we use for 
the observations in Fig.\ \ref{fig:map}).  In order to produce this
map, we have smoothed the most luminous star
particles to reduce noise (although some noise in the maps is 
visible)\footnote{In detail, we split star particles
more luminous than 8 solar luminosities (sufficient to 
produce more than 1 MSTO star; \citealp{bell08})
into $L_r/8L_{\sun}$ particles, where $L_r$ is the luminosity
of the star particle of interest.  In order to determine the placement
of the new particles in 3D space, we find the nearest 
particles (between 100 and 1000 particles, depending on luminosity)
drawn from the same satellite as the particle of interest (in order
to preserve stream/satellite morphology)
and distribute the new star particles in a space defined 
by these nearest neighbors.  In practice, the details
of this procedure are not critical for this application.}.
The maps are luminosity-weighted (intensity scale), and 
color-coded by stellar metallicity  (blue shows ${\rm [Fe/H]} = -1.5$,
red shows ${\rm [Fe/H]} = -0.5$).    
The three halos were chosen to have differing amounts of 
substructure; model 4 (left) has a modest amount of substructure and model 1
(right) has prominent substructure, especially in the most distant bins.
The key point to take away from Fig.\ \ref{fig:theorymap} is that 
it clearly demonstrates the prediction that different 
substructures should have 
distinctive stellar populations (in this case 
parameterized by metallicity). This is in {\it qualitative} agreement
with the observations --- a quantitative match will be impossible 
to produce until the conditions under which BHB stars form 
are substantially better understood.

\section{Discussion}
\label{sec:disc}

We have presented a 
map of spatial stellar population variations in the Milky Way's halo, using the 
ratio of BHB to MSTO stars.  This map is reminiscent of analogous maps for 
M31 \citep{mcconnachie09} and shows that descriptions of stellar halo populations in terms of radial gradients alone is an incomplete description of the data; much of the population variation is stream-like in morphology.  

Our results may bear on the debate about the 
controversial nature and origin of the low-latitude structure: 
it may represent debris 
from a disrupted dwarf galaxy \citep[e.g.,][]{pena05}, or 
primarily consist of material stirred up from the outer 
disk of the Milky Way \citep[see, e.g.,][for 
discussion of this possibility]{ibata05,momany06,kaz08}.
Fig.\ \ref{fig:map} shows the low-latitude 
structure to be almost devoid of BHB stars, 
implying in a broad sense a lack
of metal-poor old populations.  This is in contrast
to many of the other halo substructures (at least parts of Sagittarius, 
Virgo, and Hercules/Aquila).  
If it is found in the future that the outer thin disk of the Milky
Way is deficient in BHB stars, this may provide tentative
support to the notion that the 
low-latitude structure is composed primarily of material stirred up from
off the Milky Way's outer disk.  At the very least, the 
lack of BHB stars (one of the few options for 
precise distance determination) 
is a significant practical challenge for those attempting 
to understand the 3-dimensional distribution of the low-latitude 
structure. 

A feature of particular interest in Fig.\ \ref{fig:map} is 
the change in BHB content along the 
Sagittarius tidal stream.  It is interesting to note that 
\citet{niederste} argue for a constant BHB/MSTO ratio along
the Sgr stream in the very region that we claim a significant difference
in BHB/MSTO ratio (from $\sim 1/55$ to $\sim 1/80$).  \citet{niederste} 
subtract off the CMD of the region with the same 
$b$ but $l_{\rm control} = 180-l_{\rm Sgr}$, then count 
the number of MSTO stars and BHB stars left in the 
residual CMD.  We have confirmed that the BHB/MSTO
ratios of the control areas systematically change, from $\sim 1/50$ in 
the mirror region of the BHB star-rich part of Sgr, to $\la 1/250$ in 
the mirror region of the BHB star-poor part of Sgr (this is apparent in 
Fig.\ \ref{fig:map}; the top part of the last set of panels is relatively
BHB-rich, whereas the lower parts of the last set of panels is poor in BHB stars
both in Sgr and elsewhere).  This change in 
the properties of the control sample drives the apparent constancy
of the BHB/MSTO ratio in their work.  It is not obvious (to us at least)
how to properly interpret such a situation.  From the perspective of an
underlying smooth stellar halo (with abrupt changes in BHB/MSTO in 
the `smooth' halo) with a superimposed stream, subtracting off the control
fields is defensible.  Yet, from the perspective of viewing the stellar
halo as a combination of a number of structures in various stages
of disruption, subtraction of a control field is less defensible, and 
would lead to artificial changes in the inferred properties of the 
structure of interest.

Viewed from the latter perspective, such a
difference in the BHB/MSTO ratio would demonstrate that population 
gradients within a progenitor galaxy could in practice lead to 
population differences in the resulting tidal debris (a gradient
in the Sagittarius tidal tail has been detected before using other 
stellar population diagnostics; e.g., \citealp{martinez04,bellazzini06,chou07}).
It may well be that the apparent abruptness of the change
in the BHB/MSTO ratio between the Sagittarius debris in the anticenter
direction and the NGP direction is because the 
debris streams in those directions were stripped from Sagittarius 
during different passages (see, e.g., \citealp{law05} and \citealp{law10} 
for models in which these two parts of the debris stream were stripped
at significantly different times).  
Material closer to the edges of a satellite is preferentially stripped first;
thus, in this interpretation we would hypothesize that the 
outermost parts of the Sagittarius dwarf galaxy were devoid
of BHB stars, whereas the central parts of the dwarf were richer
in BHB stars\footnote{It is worth noting that this picture may run 
counter to the scenario suggested by the observed differences 
in e.g., BHB to red clump star ratio, or metallicity, between 
the Sagittarius core and the tidal streams \citep{bellazzini06,chou07}.  In their picture
the core of Sagittarius is more metal-rich, and the tidal tail
more metal poor.  Here, we are sensitive to differences between different
parts of the tidal tail; how that relates to earlier core/tail comparisons 
is not entirely clear.}.  Spectroscopic follow-up of stream members, 
with the goal of weeding out phase-mixed stars from stars in a dynamically-cold
stream, will help to differentiate between the two possible interpretations 
of the apparent population gradient in Sgr (see \citealp{keller10} for 
such an investigation of the Sgr trailing arm).

\section{Conclusions}
\label{sec:conc}

In this paper, we have studied the spatial structure of
stellar population variations in 
the stellar halo of the Milky Way.   We made use of 
a new color selection method
to isolate a sample of high-probability BHB star candidates,
and compare their spatial density to that of color-selected MSTO stars (taken 
to represent the general stellar content of the halo).  The abundance
of BHB stars (vs.\ MSTO stars) is known to vary strongly among 
stellar populations with globular cluster-like ranges 
in age and metal abundance, where broadly speaking high 
BHB abundance signposts particularly
old and metal-poor populations, and redder Horizontal Branch populations
are characteristic of younger, more metal-rich populations,
though the physical origin of such variations
is currently debated.  
We mapped the relative distributions of BHB and MSTO stars 
across the Heliocentric distance range 
$5 \la r/{\rm kpc} \la 30$ for $\sim 1/4$ of the celestial sphere, 
providing a panoramic view of the content
of the stellar halo.  

We found
large variations of the BHB/MSTO star ratio in the stellar halo.  Most 
importantly, variations trace different previously-identified structures, 
indicating distinct populations and hence origins for them (in common, for
example, with M31; \citealp{mcconnachie09}).  
Some halo features, e.g., 
the low-latitude structure, appear to be almost 
completely devoid of BHB stars, whereas other 
structures appear to be rich in BHB stars.  The 
Sagittarius tidal stream shows an apparent variation in the BHB/MSTO 
ratio along its extent, which we interpret in terms of 
population gradients within the progenitor dwarf galaxy
leaving observable signatures in our stellar halo.
In a previous paper \citep{bell08}, we had shown that the level of density
substructure in the Milky Way's stellar halo is consistent 
with models \citep[e.g.,][]{bullock05} in which this component
is built up exclusively from disrupted satellites.  In this paper
we have shown that another prediction of such models 
is qualitatively borne out: significant population variations, traced
by the BHB/MSTO star ratio in the Milky Way's stellar halo, and with 
a spatial structure that correlates with the density substructures.
This lends further observational support to the view that the 
stellar halo is predominantly assembled from the disrupted
debris of dwarf galaxies.

\acknowledgements
We thank the referee for their helpful suggestions.
We wish to thank Jorge Penarrubia and Frank van den Bosch for 
useful discussions.
C.\ R.\ was
supported by the Emmy Noether Programme of the Deutsche
Forschungsgemeinschaft, and is a member of the 
Heidelberg International Max Planck Research School program.  
D.\ W.\ H.\ is a research fellow of the 
Alexander von Humboldt Foundation of Germany.

Funding for the SDSS has been provided by the Alfred P. Sloan Foundation, the Participating Institutions, the National Aeronautics and Space Administration, the National Science Foundation, the U.S. Department of Energy, the Japanese Monbukagakusho, and the Max Planck Society. The SDSS Web site is http://www.sdss.org/.

The SDSS is managed by the Astrophysical Research Consortium (ARC) for the Participating Institutions. The Participating Institutions are The University of Chicago, Fermilab, the Institute for Advanced Study, the Japan Participation Group, The Johns Hopkins University, Los Alamos National Laboratory, the Max-Planck-Institute for Astronomy (MPIA), the Max-Planck-Institute for Astrophysics (MPA), New Mexico State University, University of Pittsburgh, Princeton University, the United States Naval Observatory, and the University of Washington.

\end{document}